# Do people apply different norms to humans and large language models acting on their behalf? Evidence from norm elicitations in two canonical economic games

Paweł Niszczota [a,c,*] and Elia Antoniou [b,a,c]

*Accepted for publication in Journal of Experimental Social Psychology, see*

https://doi.org/10.1016/j.jesp.2026.104991

[a] Poznań University of Economics and Business, Humans & Artificial Intelligence Laboratory (HAI Lab), Institute of International Business and Economics, al. Niepodległości 10, 61-875 Poznań, Poland; pawel.niszczota@ue.poznan.pl

[b] University of Cyprus, 1 Panepistimiou Avenue, 2109, Aglantzia, Nicosia, Cyprus; antoniou.elia@ucy.ac.cy

[c] Equal contributions.

* Corresponding author: Paweł Niszczota, Poznań University of Economics and Business, al. Niepodległości 10, 61-875, Poznań, Poland; telephone number: +48 61 854 33 24; pawel.niszczota@ue.poznan.pl; ORCID: 0000-0002-4150-3646

## Declarations

**Funding.** This research was supported by grant 2021/42/E/HS4/00289 from the National Science Centre, Poland.

**Competing interests.** The authors have no relevant financial or non-financial interests to disclose.

**Ethics approval.** Approval was obtained from the Committee of Ethical Science Research at the Poznań University of Economics and Business (resolution 1/2023). The procedures used in this study adhere to the tenets of the Declaration of Helsinki.

**Consent.** Informed consent was obtained from all individual participants included in the study.

**Data, materials, and code availability.** The pre-registration documents, data, materials, and code are available at: https://osf.io/2yt9e.

**Declaration of generative AI and AI-assisted technologies in the manuscript preparation process.** During the preparation of this work, the authors used OpenAI GPT-5.6 Sol Pro and Anthropic Claude Fable 5 to assist with proofreading, consistency checks, and language and code clarity. After using this tool, the authors reviewed and edited the content as needed and take full responsibility for the content of the published article.

### CRediT authorship contribution statement

**Paweł Niszczota:** Conceptualization, Data curation, Formal analysis, Funding acquisition, Investigation, Methodology, Project administration, Resources, Software, Supervision, Validation, Visualization, Writing – original draft, Writing – review & editing.

**Elia Antoniou:** Conceptualization, Formal analysis, Investigation, Methodology, Software, Validation, Writing – original draft, Writing – review & editing.

**Short title**: Do people apply different norms to humans and LLMs acting on their behalf?

**Word count**: 12,016 words

# Do people apply different norms to humans and large language models acting on their behalf? Evidence from norm elicitations in two canonical economic games

## Abstract

While delegating tasks to large language models (LLMs) can save people time, there is growing evidence that offloading tasks to such models produces social costs. We elicit appropriateness of behavior in two canonical economic games to study whether people have different norms for humans and LLMs acting on their behalf. More specifically, we study the social appropriateness of a spectrum of possible behaviors: when LLMs divide resources on our behalf (Dictator Game and Ultimatum Game) and when they evaluate the fairness of resource splits (Ultimatum Game). We use the Krupka-Weber norm elicitation task to detect possible shifts in social appropriateness ratings. Results of two pre-registered and incentivized experimental studies using representative samples from the UK and US ($N$ = 2,658) show three key findings. First, in the Dictator Game, the most socially appropriate allocation remains the equal split when the Dictator is a human or an LLM acting on behalf of a human, but self-serving allocations are judged less appropriate when implemented by an LLM. Second, in the Ultimatum Game, rejecting offers from an LLM Proposer is judged as more appropriate than rejecting identical offers from a human Proposer, with this difference concentrated among relatively generous offers. Lastly, rejection decisions implemented by an LLM Responder are judged no less appropriate than identical rejection decisions implemented by a human Responder. *Post hoc*, these patterns are consistent with the notion that people grant LLMs bounded agency in the investigated resource-allocation scenarios. Our findings have implications for the design of AI systems.

# Do people apply different norms to humans and large language models acting on their behalf? Evidence from norm elicitations in two canonical economic games

## Introduction

Advances in natural language processing and the widespread availability of Large Language Models (LLMs) such as GPT, Claude, Gemini, and DeepSeek make it increasingly feasible for people to delegate tasks to artificial intelligence (AI). Existing research shows that people often behave differently when interacting with AI systems than when interacting with humans (Barak & Costa–Gomes, 2025; Crandall et al., 2018; Dvorak et al., 2025; Ishowo–Oloko et al., 2019; Stinkeste et al., 2025). Across a range of economic games, human-AI interactions have been associated with lower levels of cooperation, trust, fairness, reciprocity, and moral emotions compared to human-human interactions. For example, Dvorak et al. (2025) report lower fairness, trust, trustworthiness, cooperation, and coordination when humans interact with ChatGPT rather than with another human across multiple two-player economic games. Similarly, Borthakur et al. (2025), using the Ultimatum Game, find that participants are more likely to reject disadvantageous offers from an AI than from a human, but less likely to reject advantageous offers from an AI than from a human.

What remains unclear is whether, in addition to the aforementioned difference in behaviors, there is also a shift in how the machines *ought* to behave. Injunctive norms, or what "ought to be done", reflect perceptions of which behaviors most people approve or disapprove of (Cialdini et al., 1991). Uncovering whether injunctive norms are the same for humans and LLMs acting on their behalf is a worthwhile endeavor: it helps inform the designers of AI systems that play an increasingly important role in society. AI systems now make more and more socially consequential decisions on behalf of individuals (Basu, 2026) or organizations, such as generating settlement offers or screening claims (Dong et al., 2025; Rucker et al., 2023). If, however, such actions are evaluated more strictly when implemented by an LLM working on behalf of a human rather than directly by a human, this might reduce the likelihood of humans using them. Machines have great potential to alleviate economic frictions by taking over mundane tasks (Hadfield & Koh, 2025; Risko & Gilbert, 2016), and thus, by extension, anticipating societal barriers to their adoption is of great importance. While there can be a divide between what people consider appropriate behavior and how they behave in aggregate, norms nonetheless can serve as blueprints for behavior, albeit more for some individuals than others (Fehr & Schurtenberger, 2018; Krupka & Weber, 2013).

We contribute to a strand of literature that elicits human judgments regarding how advanced technological systems should behave (e.g., Awad et al., 2018; Bonnefon et al., 2016). This perspective is particularly relevant given that the development and alignment of AI systems often involve design decisions by developers (e.g., Hagendorff, 2020; Huang et al., 2024). Our norm-elicitation approach also contributes to a growing body of work seeking to better understand how AI systems can be aligned with human values and norms (e.g., Gabriel, 2020; Koster et al., 2022).

### Present work

To shed light on whether injunctive norms differ in select decisions related to resource allocation, we use the Krupka-Weber norm-elicitation method (Krupka & Weber, 2013) for two canonical economic games: the Dictator Game and the Ultimatum Game. In Experiment 1, we elicit norms concerning the appropriateness of dividing resources in a Dictator Game. In Experiment 2, we elicit norms concerning the appropriateness of rejecting a previously made offer. This, in turn, corresponds to cases where we delegate to an LLM not the role of the agent making the offer, but the role of an enforcer, who has the power to terminate a potential economic transaction if it is deemed unsatisfactory.

## Theoretical underpinnings

Social norms are shared beliefs about which actions are socially appropriate in a given context (Bicchieri, 2006). In the present study, we focus on eliciting injunctive social norms. Below, we present work that points to why normative evaluations may differ across humans and LLMs.

The Computers Are Social Actors (CASA) paradigm suggests that people often apply the same social rules and expectations to computers that they apply to other people, even when only minimal social cues are present, such as language or interactivity (Nass & Moon, 2000). However, Gambino et al. (2020) argue that repeated interaction with technology has also led people to develop distinct scripts for engaging with computers that differ from those used in human-human interaction. Building on this perspective, in this paper we evaluate whether normative evaluations differ for humans and LLMs, and develop hypotheses based on established theoretical accounts.

Research in social cognition shows that people evaluate behavior not only in terms of the action itself or its consequences, but also in terms of why the action was performed (Heider, 1958; Jones & Davis, 1965; Reeder, 2009). People therefore infer motives from behavior, and these inferences shape their judgments (Reeder, 2009). The same action, for example, may be evaluated differently depending on whether it is seen as motivated by self-interest, fairness, or some other concern. Importantly, such inferred motives are not directly observable. Rather, they are constructed from cues, including characteristics of the actor and features of the situation (Carlson et al., 2022). When a human delegates a decision to an LLM, observers may infer that the person strategically chose delegation to produce a desired outcome while avoiding direct responsibility for it (Bartling & Fischbacher, 2012; Köbis et al., 2021). Consequently, evaluations of actions implemented by an LLM may reflect not only the decision outcome itself but also the motives observers attribute to the decision to delegate.

These inferences may also depend on the role assigned to the decision-maker. Social role theory suggests that expectations about appropriate behavior depend not only on who or what the actor is, but also on the role the actor occupies (Biddle, 1986; Eagly & Wood, 2012). When an LLM is described as acting on behalf of a human, it may be perceived less as an autonomous actor and more as that person's instrument. This framing may activate expectations similar to those found in principal-agent relationships, in which agents are expected to act in the principal's interest (Eisenhardt, 1989). Under these conditions, observers may attribute the motives underlying the action primarily to the human principal rather than to the LLM itself. In turn, they may expect the LLM to advance the human's interests rather than any "interests" of its own.

Together, these perspectives suggest that normative evaluations may be shaped by the motives observers attribute to the human principal, as well as by beliefs about which roles machines can appropriately occupy in resource-allocation contexts. This argument also connects to mind perception theory (Gray et al., 2007). As LLMs capabilities increase, people may perceive them as having greater agency, consistent with recent evidence (Jacobs et al., 2024; Lee & Kim, 2025). Yet it remains unclear whether greater perceived agency is sufficient to legitimize machines as actors who make or enforce resource-allocation decisions. If normative evaluations of machines acting on behalf of humans differ from evaluations of humans making the same decisions, this would indicate that the type of the agent matters for injunctive evaluations. If they do not differ, this would suggest that the type of the agent does not affect injunctive evaluations.

Our study does not directly measure the mechanisms discussed above, such as inferred motives, perceptions of agency, or social role. These mechanisms should therefore be understood as theoretically grounded motivations of our work.

## General Methods

### Economic games

To examine whether normative expectations differ for humans and LLMs acting on their behalf, we use two canonical economic games: the Dictator Game and the Ultimatum Game, which relate to fairness norms and norm enforcement in economic interactions. By contrast, cooperative games such as the Prisoner's Dilemma or the Trust Game involve strategic interdependence and expectations about others' future behavior, which makes it more difficult to isolate normative evaluations of specific actions.

#### *Dictator Game*

The Dictator Game (Kahneman et al., 1986) is a canonical economic game, involving a single decision: the "Dictator" has to decide how to split money between herself and a Recipient. This game is often interpreted as a measure of prosociality, as the Recipient has no way to reciprocate an unfavorable split made by the Dictator. Under standard economic assumptions, a purely self-interested Dictator would give nothing to the Recipient, corresponding to the equilibrium prediction. However, empirical evidence shows that people rarely do so (Kahneman et al., 1986; Levitt & List, 2007). This has been attributed to the psychological cost of deviating from the social norm of an equal split between the Dictator and the Recipient (Kimbrough & Vostroknutov, 2016).

Since the game does not allow the Recipient to respond, we elicit social appropriateness ratings of the different allocations. This is in contrast to the Ultimatum Game, where the second player can reject offers deemed inappropriate (Güth et al., 1982).

#### *Ultimatum Game*

The Ultimatum Game was first introduced by Güth, Schmittberger, and Schwarze (1982) to study bargaining behavior and fairness in economic exchanges. In this game, one player (the Proposer) is given a sum of money and must decide how to divide it between themselves and another player (the Responder). The Responder can either accept or reject the offer. If the Responder accepts the offer, both players receive the proposed amounts. If the Responder rejects the offer, both players receive a payoff of zero.

Predictions under rational choice theory posit that any offer higher than zero should be accepted (Fehr & Schmidt, 1999). However, experimental results show that responders often reject offers they perceive as unfair (Oosterbeek et al., 2004). This finding highlights the importance of both fairness considerations and social norms in economic decision-making. Because fairness considerations and social norms drive rejections, proposers anticipate them and often make strategically higher offers to increase the likelihood of acceptance.

**Norm elicitation procedure**

To elicit norms concerning appropriate social behavior, we used the Krupka-Weber norm elicitation procedure (Krupka & Weber, 2013). Specifically, for the Dictator Game, participants were presented with 11 different possible allocation decisions (e.g., "The Dictator gives £0 to the Recipient", "The Dictator gives £1 to the Recipient", etc.) and were asked to rate how socially appropriate the action was on a four-point scale ranging from "very socially inappropriate" to "very socially appropriate".

Participants were informed that one of the 11 allocation decisions would be randomly selected for a bonus payment. They received the bonus payment if their rating for that split matched the modal response of other participants for the specific split within the same treatment. Thus, participants had a financial incentive to report what they believed most people would consider socially appropriate for all splits.

For example, suppose that the action "Give £5 to Participant B (Participant A gets £5, Participant B gets £5)" is randomly selected for the bonus payment. If a participant rates this action as "very socially appropriate" and this rating is the modal rating among participants in the same treatment, then the participant will receive the bonus payment. If it is not the modal rating, then the participant will not receive the bonus payment. A similar procedure was used in the Ultimatum Game, with participants evaluating the social appropriateness of each possible rejection decision. This procedure has been shown to be robust, and appropriate for use on crowdsourcing platforms (König–Kersting, 2024).

Under the Krupka-Weber procedure, social norms correspond to the appropriateness of the full set of possible actions. In the Dictator Game, equal splits typically receive the highest level of social approval, although there are cases in which this evaluation can shift (Kimbrough & Vostroknutov, 2016). In the Ultimatum Game, fair offers are generally evaluated as socially appropriate, and rejecting unfair offers is often viewed as norm-consistent, even when personally costly (Güth et al., 1982).

**Adherence to good practices**

In both experimental studies, we report all measures, manipulations, and exclusions. The sample sizes were determined before any data analysis. Sample sizes were determined prior to data collection based on a priori power simulations. No deception was used in either experiment.

## Experiment 1

**Hypotheses**

We use the Dictator Game to study normative evaluations in a setting where the decision-maker allocates resources independently and the recipient cannot respond. This feature of the game is theoretically important because there is no strategic incentive for an LLM, acting on behalf of a human, to offer a fair split to the recipient. Therefore, evaluations of allocations will reflect normative judgments rather than strategic considerations.

In this setting, when an LLM acts on behalf of a human, evaluations may be shaped by how observers infer the motives behind the decision, consistent with research in social cognition (Heider, 1958; Jones & Davis, 1965; Reeder, 2009). In particular, the use of an LLM may lead participants to attribute the resulting allocation not to the LLM itself, but rather to the human behind it. As a result, self-serving allocations implemented by an LLM, acting on behalf of a human, may be perceived as less socially appropriate than identical allocations made directly by a human.

This reasoning is also consistent with the mind perception theory (Gray et al., 2007), which suggests that machines are perceived as having some agency, but very little capacity for experience, such as intentions, needs, or feelings. We posit that since the LLM is not perceived as having needs, self-serving allocations implemented by the LLM may be less justified and thus more likely to be attributed to the intentions of the human behind it.

This reasoning suggests that participants may apply a stricter normative standard to allocations implemented by the LLM. This is particularly the case for allocations that favor the human behind the machine, as in such circumstances observers may be more likely to interpret the delegation as strategically motivated or self-serving. By testing both the most appropriate allocation and the full distribution of evaluations, we can identify where differences between human and LLM evaluations emerge.

Hypothesis 1 posits a shift in normative evaluations, such that the most socially appropriate allocation (i.e., the peak of the norm profile) is higher when implemented by an LLM acting on behalf of a human than when implemented directly by a human. In the Dictator Game, the equal split is commonly interpreted as the fairness benchmark (Engel, 2011; Kahneman et al., 1986), thus representing the focal allocation. Our first hypothesis, partially motivated by the empirical findings of Borthakur et al. (2025), posits a stricter normative standard and tests whether there is a shift in what is perceived as the most socially appropriate allocation.

**Hypothesis 1:** The most socially appropriate split will be higher when the Dictator is an LLM (acting on behalf of a human) rather than a human.

Hypothesis 2 represents a weaker version of Hypothesis 1, suggesting that differences in evaluations may emerge at the level of the full distribution of allocation splits. Under the *inferred motives* interpretation, observers may overall attribute such allocations to intentional self-interest on the part of the human behind the LLM. As a result, these allocations are expected to be evaluated as less socially appropriate, even if the most socially appropriate allocation (i.e., the peak of the norm profile) remains unchanged.

**Hypothesis 2:** Social appropriateness ratings will be lower when the Dictator is an LLM (acting on behalf of a human) instead of a human.

For Hypothesis 3, we build on prior evidence that fairness norms in the Dictator Game are asymmetric (Fehr & Schmidt, 1999; Kimbrough & Vostroknutov, 2016). Specifically, allocations below the equal split may be seen as violations of distributive fairness, whereas allocations above the equal split are typically viewed as generous (Gächter et al., 2017; Krupka & Weber, 2013). Under the *inferred motives* account, allocations below the fairness benchmark may be evaluated as less appropriate when implemented by an LLM, as such allocations may be interpreted as reflecting intentional self-interest on the part of the human behind the LLM. This reasoning is also consistent with mind perception theory (Gray et al., 2007), which posits that machines are not perceived as having needs or interests. We therefore expect that differences in evaluations between LLM and human Dictators will be stronger for splits that favor the Dictator than for those favoring the Recipient.

**Hypothesis 3:** The effect of the Dictator being an LLM (acting on behalf of a human) and not a human will be more adverse for splits in which the Dictator gets more, than in splits in which the Recipient gets more.

To sum up, a social norm as defined by Krupka and Weber (2013) is the full profile of socially appropriate and inappropriate evaluations across all available actions. Our hypotheses test different dimensions of potential differences in this profile. Specifically, Hypothesis 1 examines differences in the most socially appropriate allocation (i.e., the peak of the norm profile), Hypothesis 2 examines changes in the overall distribution, and Hypothesis 3 examines asymmetries in evaluations around the fairness benchmark.

## Methodology

### *Experimental design*

The experiment had a mixed design since each participant rated all 11 splits. The between-subjects factor was the Dictator type, who was either a human or an LLM acting on behalf of a human (who delegated this task to the LLM). Each participant rated the social appropriateness of 11 ways of splitting £10 in the Dictator Game. Participants completed the Krupka-Weber norm elicitation task (Krupka & Weber, 2013), in which they were incentivized to correctly answer what most people think about a particular split. The experiment was programmed with oTree (Chen et al., 2016).

### *Participants*

We recruited a representative sample of people from the United Kingdom through Prolific (using the "Representative sample" option on that platform, which ensured that the sample was representative of the UK population in terms of sex, age, and ethnicity based on 2021/22 census data). We recruited 580 people in each of the two between-subjects conditions, resulting in a final sample of 1160 people. The sample size was based on a power simulation, which was intended to obtain 80% statistical power at $\alpha = 5\%$ to detect a small effect in the between-subjects condition (defined as Cohen's $d = 0.20$; Cohen, 1988). Given that there are 11 possible splits in the Dictator Game and thus 11 tests, we used the Holm-Bonferroni method to counter increased Type I error rates. Power was estimated by Monte Carlo simulation. In each simulated dataset, repeated ordinal responses on the four-category scale were generated from prespecified category probabilities, with dependence between responses from the same participant, and analyzed using a two-sided clustered Wilcoxon rank-sum test at each of the 11 allocation levels. This mirrors the analytical approach of the foundational Krupka-Weber study and provides a conservative, tractable basis for sample-size planning. Our analysis showed that 550 participants per condition would ensure >80% power after the Holm-Bonferroni correction; the code used to perform the power analysis is available at OSF. We recruited more participants per condition to account for a ~5% exclusion rate of inattentive participants. We excluded 7 participants (using one attention check instead of two, which we will discuss later) who did not answer an attention check concerning what amount was split in the game, leaving a final sample of 1153 participants ($M_{age} = 47.0$, $SD = 15.1$; 50.04% female, 48.92% male, 1.04% non-binary or undisclosed). There were 576 participants in the *Human Dictator* condition, and 577 participants in the *LLM Dictator* condition. See **Table S1** in the Supplementary Materials for the descriptive statistics.

Participants were paid a flat fee of £0.75 and told that they could earn an additional £0.50 if they successfully identified the most common answer in a randomly selected choice from the sample. All sessions took place in September 2025.

### *Statistical analysis*

The outcome variable is the perceived social appropriateness of different monetary offers in the Dictator Game. Participants rated how socially appropriate each possible offer (e.g., 0% to 100% of the endowment) was on a four-point scale ranging from "Very Socially Inappropriate" to "Very Socially Appropriate". Following the standard scoring convention from Krupka and Weber (Krupka & Weber, 2013), we mapped the four categories onto evenly spaced values on the [–1, 1] interval. Specifically, the ratings "Very Socially Inappropriate", "Somewhat Socially Inappropriate", "Somewhat Socially Appropriate", and "Very Socially Appropriate" were assigned the values –1, –1/3, 1/3, and 1, respectively. This transformation converts the ordinal categories into an evenly spaced numerical scale for analysis.

The key independent variable is the type of Dictator, manipulated between-subjects with two levels: Human and LLM (after a human delegated the decision to the LLM). In the Human condition, participants were told that Participant A was randomly paired with another person, Participant B. In the LLM condition they were told that Participant A had delegated the task to an LLM and the LLM was then randomly paired with Participant B.

In addition, we also collected basic demographic variables, including age, gender, education level, occupation, field of study, income level, trust in LLMs, frequency of AI or large language model use, and prior experience with economic games. These variables were used for robustness checks (refer to the Supplementary Materials). Consistent with Krupka and Weber (2013), we relied on non-parametric tests to determine whether social appropriateness ratings differ between conditions.

We tested Hypothesis 1 via a participant-clustered bootstrap test (10,000 replications) aimed at testing the null hypothesis that the most socially appropriate split (i.e., the peak of the norm profile) is the same when the Dictator is an LLM rather than a human. Within each between-subjects condition, we selected the split with the highest mean appropriateness. Hypothesis 1 would be supported if the two-sided 95% CI for $\Delta_{peak}$ = *argmax Social appropriateness(split in LLM condition) – argmax Social appropriateness (split in Human condition)* excludes zero. To complement our primary analysis, we computed a series of Wilcoxon tests on each of the splits.

We tested Hypothesis 2 via a clustered Wilcoxon rank-sum test (using the *clusrank* package in *R*; Jiang et al., 2020), testing the null hypothesis that there is no difference in ratings in the human and LLM condition (two-sided test at $\alpha = 0.05$). This test accounts for the fact that each participant rates each of the 11 splits (i.e., data points are not independent). Given that we are also interested in which splits there is a difference in human/LLM Dictator ratings, we also conducted 11 Wilcoxon tests, which were Holm-Bonferroni-corrected.

We tested Hypothesis 3 via a participant-clustered bootstrap test (10,000 replications) testing the null hypothesis that the difference in ratings for LLM and human dictators is the same in “low” splits, that favor the Dictator (Recipient receives £0-4) and “high” splits, which favor the Recipient (Recipient receives £6-10). We computed Cliff’s Δ from the Mann-Whitney-Wilcoxon (see **Supplementary Materials** for details) and considered the effects to differ between low and high splits if the two-sided 95% CI for $\Delta_{high–low} = \Delta_{high} – \Delta_{low}$ excludes zero. Note that despite the directional nature of the hypotheses, we used two-sided confidence intervals/tests in these instances. We did so to be conservative and remain open to the possibility of an effect in the opposite direction (Lombardi & Hurlbert, 2009).

***Open science practices***

The experiment was pre-registered on the Open Science Framework at https://osf.io/n5wq4/overview?view_only=bd37c0fffcd94d80a16627007338ce7b. Data, code and materials are available at https://osf.io/2yt9e/overview?view_only=8268c7ac95d94f2390fd3ed6dc76269a. The only deviation from the pre-registration – made to make fuller use of the representative nature of the sample – is that we decided to exclude participants based on one exclusion criterion and not two. A substantial portion of participants failed one of the attention checks, which is most likely attributable to an ambiguity inadvertently introduced by us (i.e., participants may have misunderstood this attention check). However, results are qualitatively the same. In **Table S4** in the **Supplementary Materials** we present results using the pre-registered exclusions, as well as alternative exclusions.

## Results

### *The most socially appropriate behavior*

We first compared whether the peak of the norm profile – the most appropriate action – is the same for human Dictators and LLM Dictators (i.e., whether $\Delta = 0$). In both conditions, the most socially appropriate split was the equal split, where the Dictator and Recipient each receive £5. A bootstrapping test confirmed that this result is highly stable (95% *CI* for $\Delta = [0, 0]$). These results indicate that delegating the decision on how to split money in the Dictator Game to an LLM does not alter what is considered the most appropriate action (see **Figure 1A**). Moreover, in both treatments, the most socially inappropriate split was the allocation in which the Dictator kept the full amount (£10), leaving the Recipient with £0.

**Figure 1**

*Social appropriateness of offering a split in Dictator Game (Experiment 1) and rejecting split in Ultimatum Game (Experiment 2)*

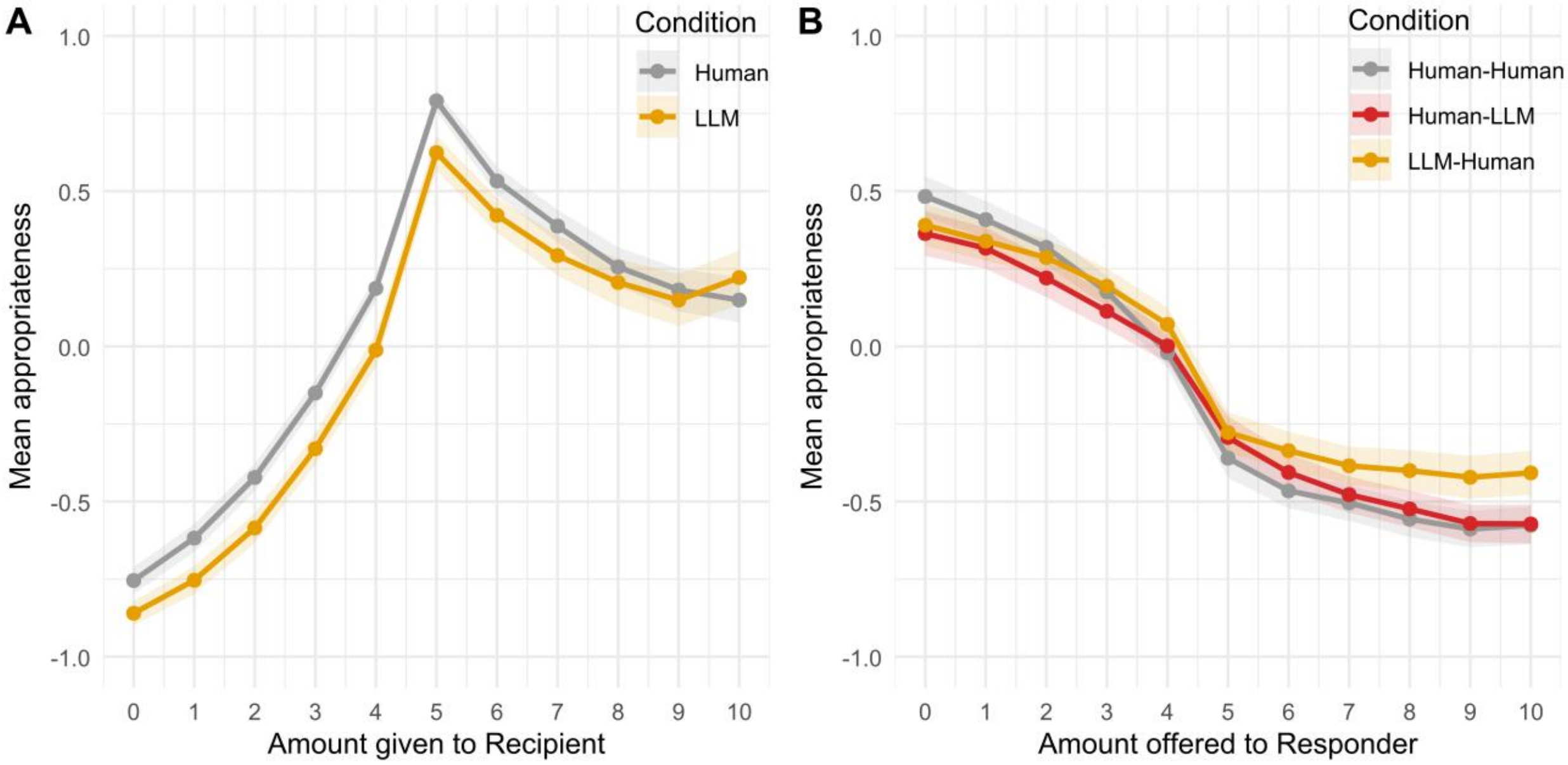

*Note. We recoded responses as follows: "Very Socially Inappropriate", "Somewhat Socially Inappropriate", "Somewhat Socially Appropriate", and "Very Socially Appropriate" were assigned the values –1, –1/3, 1/3, and 1, respectively. Shaded areas represent 95% CIs.*

Nonetheless, there was some divergence in opinions regarding the social appropriateness of various splits. **Figure 2A** shows the proportion of participants who assigned their maximal social appropriateness rating to each split. The equal split (£5-£5) received the highest approval in both conditions, with roughly 40% of participants rating it as the most socially appropriate action. There was also greater disagreement for the split in which the Dictator gives the entire amount (£0-£10), with approximately 10% of participants selecting it as the most socially appropriate in the human condition, compared to nearly 20% in the LLM condition. We also computed the Kullback-Leibler (KL) divergence between rating distributions (**Figure 2B**).

**Figure 2**

*Divergence of opinions – Experiment 1*

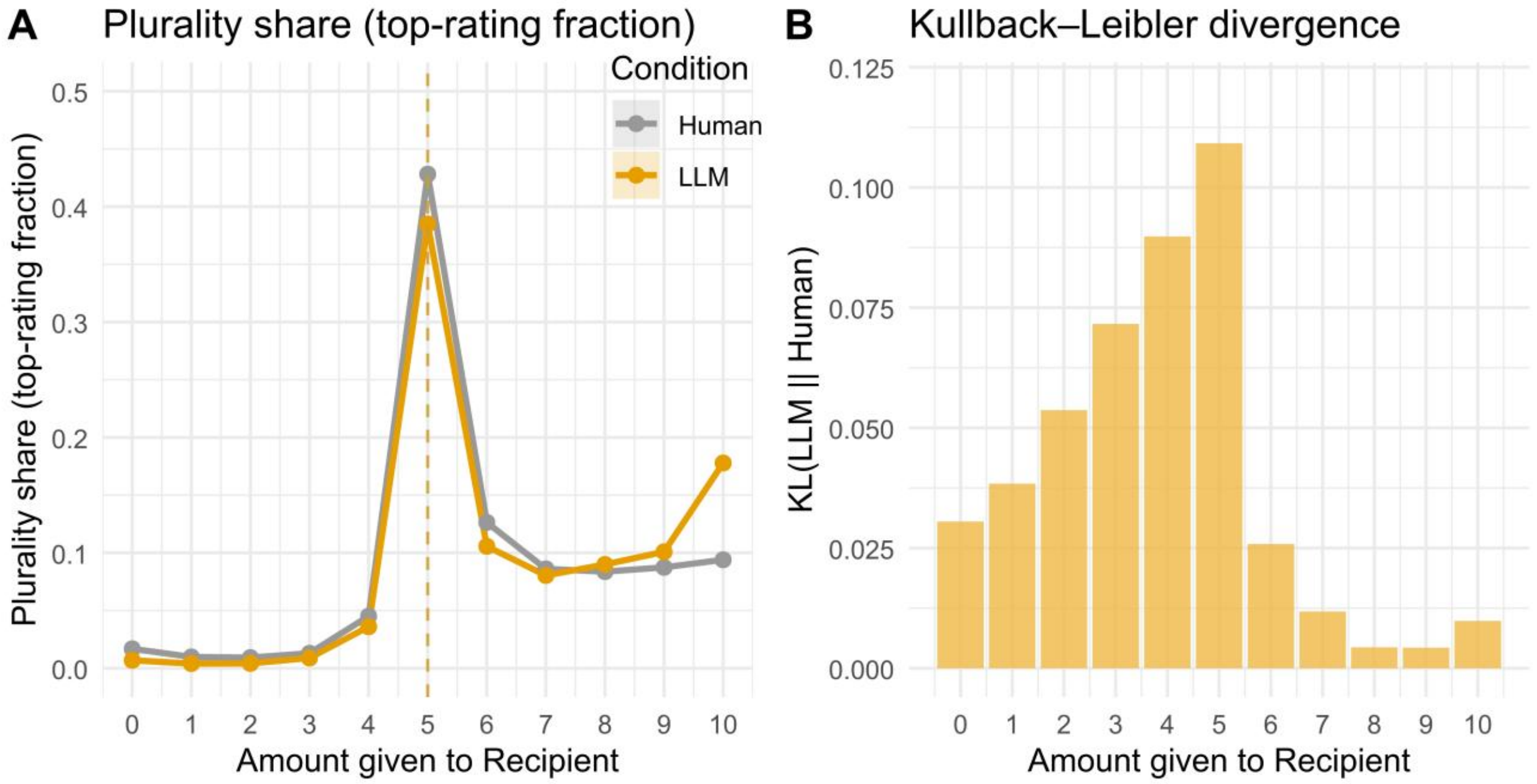

*Note. In Panel A, we show what portion of participants have given the highest social appropriateness score for a specific split. In Panel B, we report the Kullback-Leibler divergence between responses in the Human and LLM Dictator conditions. Higher scores indicate a greater level of divergence.*

***Downward shift in the appropriateness of LLM behavior***

We further performed a clustered Wilcoxon rank sum test, which indicates that, when considering appropriateness ratings across all of the splits, ratings for human and LLM Dictators differ significantly (*z*

= –5.12, $p < .001$). Consistent with Krupka and Weber (2013), we complement this omnibus test with a series of Wilcoxon rank sum tests for each of the 11 splits (see **Table 1**). The differences are concentrated among splits that favor the Dictator. Ratings for LLM Dictators are significantly lower than for human Dictators for transfers between £0 and £7, with the largest differences occurring around the equal split. For higher transfers (£8-£10), ratings do not differ significantly.

**Table 1**

*Difference in social appropriateness ratings – Experiment 1*

| *Amount given to Recipient* | *Difference in means* | *95% CI* | *W* | *p* |
|---|---|---|---|---|
| 0 | **–0.100*** ** | [–0.152, –0.048] | 182021 | <.001 |
| 1 | **–0.131*** ** | [–0.187, –0.075] | 188813 | <.001 |
| 2 | **–0.171*** ** | [–0.228, –0.114] | 196463 | <.001 |
| 3 | **–0.182*** ** | [–0.234, –0.125] | 198354 | <.001 |
| 4 | **–0.209*** ** | [–0.267, –0.149] | 200318 | <.001 |
| 5 | **–0.196*** ** | [–0.254, –0.138] | 194122 | <.001 |
| 6 | **–0.124*** ** | [–0.189, –0.059] | 186464 | <.001 |
| 7 | **–0.101*** | [–0.175, –0.026] | 180688 | .03 |
| 8 | –0.063 | [–0.151, 0.024] | 173736 | .34 |
| 9 | –0.045 | [–0.140, 0.051] | 171320 | .34 |
| 10 | 0.074 | [–0.025, 0.174] | 157751 | .34 |

*Note.* For each split we performed a Mann-Whitney-Wilcoxon test. Human Dictator is the benchmark, and thus negative values indicate that social appropriateness ratings were lower for the LLM Dictator than the Human Dictator. Given that there are 11 splits and for each of them we perform a separate test, we performed a Holm-Bonferroni correction of the *p*-values.

We next compare evaluations across "low" and "high" splits. The difference between human and LLM Dictators is substantially larger for splits that favor the Dictator than for those that favor the Recipient. Using Cliff's Δ as a measure of effect size, the difference is –0.163 for low splits (0-4) but only –0.047 for high splits (6-10). A participant-clustered bootstrap confirms that this difference is statistically significant ($\Delta_{high–low}$ = 0.116, 95% *CI* [0.048, 0.183]), indicating that the effect of LLM delegation is concentrated among splits that favor the Dictator.

***Robustness analyses and exploratory analyses***

As pre-registered, we also analyzed the data using linear mixed regressions. To aid interpretability, in **Table 2** we present results of estimated marginal means for these models (Searle et al., 1980). Full models containing both the single and interaction terms are presented in **Table S2** in the **Supplementary Materials**.

In the Supplementary Materials (**Table S3**) we present the result of further analyses that test whether our findings are similar if we control for the effect of such demographic information as age, gender, education, socioeconomic status, as well as trust in AI, the intensity of AI use in everyday life, and prior experience with the Dictator Game. In short, results are qualitatively the same after taking into account the effect of the control variables.

As a visual companion to our analyses, in **Figure 3** we present model predictions, which suggest that the model is well-fitted to the empirical data (see **Figure 1**). Finally, in **Figures S1 and S2** in the Supplementary Materials, we present how the normative evaluations differ between men and women, experience with the game, trust in AI, and intensity of AI use in real-life. The differences, while present, are not stark, suggesting that our key findings are not driven by a subset of our participants that we can differentiate using available data.

**Figure 3**

*Predicted social appropriateness ratings of splits in Dictator Game in Experiment 1 (A) and Ultimatum Game in Experiment 2 (B) using a linear mixed-model*

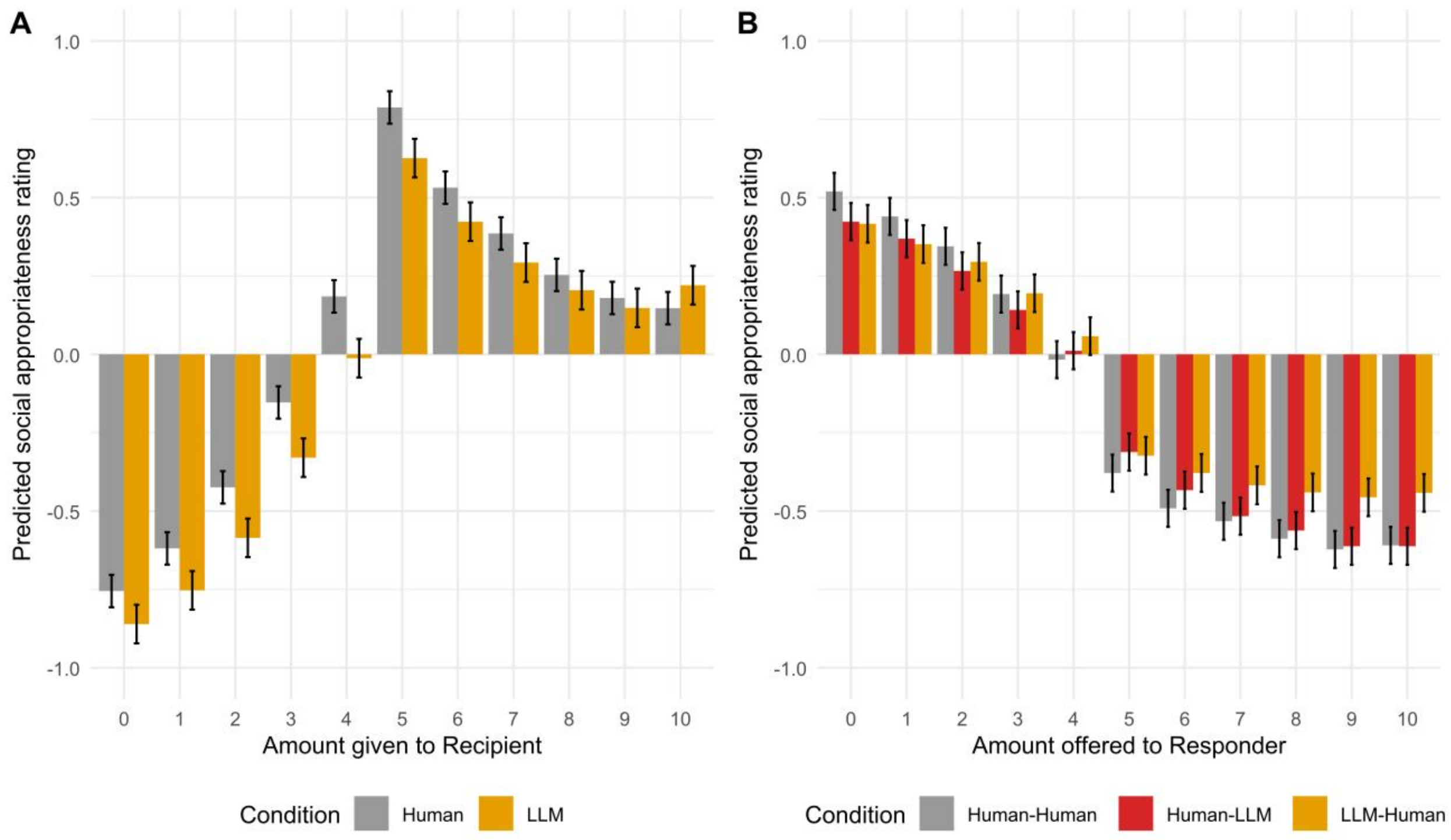

*Note. We recoded responses as follows: "Very Socially Inappropriate", "Somewhat Socially Inappropriate", "Somewhat Socially Appropriate", and "Very Socially Appropriate" were assigned the values –1, –1/3, 1/3, and 1, respectively. Bars represent 95% CIs.*

## Discussion

In Experiment 1, we examined whether decisions implemented by LLMs lead to different perceived injunctive norms in the Dictator Game. Our results showed that participants agree that the most socially appropriate action is the equal split, regardless of whether the decision is made by a human or delegated to an LLM. This suggests that participants apply the same fairness benchmark to both human and LLM Dictators, inconsistent with Hypothesis 1. In turn, the implementation of the resource allocation decision by an LLM does not cause a shift in what is considered fair: a 50–50 split remained the most socially appropriate way of splitting resources in the context of a Dictator Game (Andreoni & Bernheim, 2009).

However, social appropriateness ratings are significantly lower when allocations are implemented by an LLM rather than by a human, supporting Hypothesis 2. Moreover, our results show that allocations that favor the Dictator and were implemented by LLM Dictators were judged as less socially appropriate than identical allocations made by human Dictators. This asymmetric pattern supports Hypothesis 3.

Taken together, our findings show some consistency with the *inferred motive* account (Carlson et al., 2022), in which observers may infer that the principal had ulterior motives when delegating to a machine that behaved selfishly on the principal's behalf. While behaviors are always noisy in the sense that the very same action could be the result of different intentions (Thielmann et al., 2021), self-serving allocations implemented by an LLM may be more likely to trigger suspicions that delegation served a self-interested purpose. It is important to note that our design does not allow us to isolate whether the observed differences in evaluations arise from delegation itself, from beliefs about LLMs, or from their combination.

**Table 2**

*Estimated marginal effects in linear mixed models – Experiment 1 and Experiment 2*

| Given to Recipient (Exp. 1) or Offered to Responder (Exp. 2) | Dictator Game (Experiment 1): social appropriateness of *making* split by Dictator | | | | Ultimatum Game (Experiment 2): social appropriateness of *rejecting* split by Responder | | | | | | | |
|---|---|---|---|---|---|---|---|---|---|---|---|---|
| | LLM Dictator vs Human Dictator | | | | Human Proposer–LLM Responder vs Human Proposer–Human Responder | | | | LLM Proposer–Human Responder vs Human Proposer–Human Responder | | | |
| | *Estimate* | *95% CI* | *t* | *p* | *Estimate* | *95% CI* | *t* | *p* | *Estimate* | *95% CI* | *t* | *p* |
| 0 | **–0.100**** | [–0.171, –0.029] | –2.748 | 0.006 | **–0.106*** | [–0.189, –0.023] | –2.492 | 0.013 | **–0.086*** | [–0.170, –0.002] | –2.004 | 0.045 |
| 1 | **–0.131***** | [–0.202, –0.060] | –3.613 | <0.001 | –0.081$^{\dagger}$ | [–0.165, 0.002] | –1.919 | 0.055 | –0.072$^{\dagger}$ | [–0.156, 0.012] | –1.679 | 0.093 |
| 2 | **–0.171***** | [–0.242, –0.100] | –4.704 | <0.001 | **–0.088*** | [–0.171, –0.004] | –2.062 | 0.039 | –0.036 | [–0.120, 0.049] | –0.828 | 0.408 |
| 3 | **–0.182***** | [–0.253, –0.111] | –5.003 | <0.001 | –0.060 | [–0.143, 0.023] | –1.417 | 0.157 | 0.013 | [–0.071, 0.098] | 0.313 | 0.754 |
| 4 | **–0.209***** | [–0.280, –0.138] | –5.751 | <0.001 | 0.015 | [–0.068, 0.099] | 0.363 | 0.717 | 0.082$^{\dagger}$ | [–0.003, 0.166] | 1.898 | 0.058 |
| 5 | **–0.196***** | [–0.267, –0.125] | –5.398 | <0.001 | 0.055 | [–0.029, 0.138] | 1.286 | 0.198 | 0.064 | [–0.020, 0.148] | 1.484 | 0.138 |
| 6 | **–0.124***** | [–0.195, –0.053] | –3.413 | <0.001 | 0.047 | [–0.036, 0.131] | 1.113 | 0.266 | **0.118**** | [0.033, 0.202] | 2.739 | 0.006 |
| 7 | **–0.101**** | [–0.172, –0.029] | –2.77 | 0.006 | 0.009 | [–0.074, 0.092] | 0.212 | 0.832 | **0.119**** | [0.035, 0.204] | 2.779 | 0.005 |
| 8 | –0.063$^{\dagger}$ | [–0.135, 0.008] | –1.745 | 0.081 | 0.019 | [–0.065, 0.102] | 0.438 | 0.661 | **0.150***** | [0.066, 0.235] | 3.499 | <0.001 |
| 9 | –0.045 | [–0.116, 0.026] | –1.233 | 0.218 | 0.004 | [–0.079, 0.087] | 0.098 | 0.922 | **0.168***** | [0.084, 0.252] | 3.906 | <0.001 |
| 10 | **0.074*** | [0.003, 0.145] | 2.046 | 0.041 | –0.008 | [–0.091, 0.075] | –0.185 | 0.853 | **0.169***** | [0.085, 0.253] | 3.934 | <0.001 |
| **Random Effects** | | | | | | | | | | | | |
| $\sigma^2$ | 0.26 | | | | 0.34 | | | | 0.33 | | | |
| $\tau_{00}$ | 0.12 participant_id | | | | 0.12 participant_id | | | | 0.13 participant_id | | | |
| ICC | 0.30 | | | | 0.26 | | | | 0.28 | | | |
| N | 1153 participant_id | | | | 1001 participant_id | | | | 1003 participant_id | | | |

| | | | |
|---|---|---|---|
| Observations | 12683 | 11011 | 11033 |
| Marginal $R^2$ / Conditional $R^2$ | 0.360 / 0.555 | 0.252 / 0.444 | 0.230 / 0.445 |

*Note.* Estimates were obtained using the *lme4* (Bates et al., 2015) and *emmeans* (Lenth et al., 2025) packages in *R*. For the Dictator Game, Human is the control condition, and thus negative values indicate that social appropriateness ratings were lower in the LLM Dictator condition. For the Ultimatum Game, *Human-Proposer–Human-Responder* is the control condition, and thus negative values indicate that social appropriateness ratings were lower in the Human–LLM/LLM–Human pair. Given that there are 11 splits and for each of them we perform a separate test, we performed a Holm-Bonferroni correction of the *p*-values. Bold estimates are statistically significant at the 5% level or lower. See **Table S5** in the Supplementary Materials for estimates for regressions that include the effect of control variables.
$^{\dagger}$ $p < .10$, * $p < .05$, ** $p < .01$, *** $p < .001$.

## Experiment 2

### Hypotheses

We use the Ultimatum Game to study perceived appropriateness in a setting where both players play an active role. In contrast to the Dictator Game, the Ultimatum Game introduces a strategic interdependence, as the Responder can reject offers to enforce fairness norms at a personal cost (Fehr & Schmidt, 1999). We examine two conditions, in which an LLM occupies either the role of the Responder (Human–Proposer/LLM–Responder), or the role of the Proposer (LLM–Proposer/Human–Responder). Each condition is compared against a baseline in which both roles are performed by humans (Human–Proposer/Human–Responder).

This design allows us to isolate how the identity of the agent affects the perceived social appropriateness of rejection decisions, depending on whether the LLM acts as the Responder or the Proposer. In the Ultimatum Game, rejection represents a form of costly punishment through which responders enforce fairness norms (Fehr & Schmidt, 1999). When both players are human, rejection is typically interpreted as a socially meaningful act of norm enforcement, reflecting concerns about relative shares and justice (Bolton & Ockenfels, 2000; Fehr & Schmidt, 1999).

However, when an LLM occupies the role of the Responder, this form of norm enforcement may be evaluated differently. Social role theory suggests that expectations about appropriate behavior depend on the role an agent is perceived to occupy (Biddle, 1986; Eagly & Wood, 2012). The role of enforcing fairness norms is typically associated with human agents who are perceived as capable of evaluating and responding to fairness violations.

This reasoning is also consistent with mind perception theory (Gray et al., 2007). As a result, rejection decisions implemented by an LLM may be perceived as having less social meaning than identical decisions made by a human.

Taken together, these perspectives suggest that rejection decisions implemented by an LLM may be perceived as less socially appropriate than those implemented by a human.

**Hypothesis 4:** There is an overall downward shift in appropriateness of rejection in the Human–Proposer/LLM–Responder condition (relative to the Human–Proposer/Human–Responder condition).

When the proposer is an LLM acting on behalf of a human and the responder is human, the responder's role as the enforcer of fairness norms remains unchanged. However, the perceived legitimacy of the offer itself may change. Under the inferred motives account, offers implemented by an LLM may be interpreted as reflecting the intentions of the human who delegated the decision. In particular, offers that deviate from the fairness benchmark may be more likely to be interpreted as strategically motivated or self-serving. This interpretation is also consistent with mind perception theory, where machines are not typically perceived as having their own needs or interests. Since rejection serves as a response to perceived unfairness or questionable intention, offers that are interpreted in this way may increase the perceived appropriateness of rejection. Hypothesis 5 tests whether offers implemented through an LLM are evaluated differently, leading to a higher perceived appropriateness of rejecting them.

**Hypothesis 5:** There is an overall upward shift in appropriateness of rejection in the LLM–Proposer/Human–Responder condition (relative to the Human–Proposer/Human–Responder condition).

We further examine whether these differences vary across allocation levels. Fairness norms in the Ultimatum Game are asymmetric. Specifically, low offers constitute violations of distributive fairness, whereas high offers are consistent with or exceed the fairness benchmark (Gächter et al., 2017; Krupka & Weber, 2013).

For the case in which the Responder is an LLM, if rejection is interpreted as norm enforcement, differences in perceived appropriateness may be more pronounced for low offers, where fairness violations are more salient. In such cases, evaluations of rejection may be particularly sensitive to the identity of the agent implementing the decision.

**Hypothesis 6:** The downward shift in appropriateness of rejection in the Human–Proposer/LLM–Responder condition (relative to the Human–Proposer/Human–Responder condition) is stronger in "low" splits (when 0-4 is offered to the Responder) than "high" splits (when 6-10 is offered to the Responder).

For the case in which the Proposer is an LLM, the inferred motives account suggests that low offers may be attributed to the intentions of the human behind the LLM. Low offers may therefore be more strongly associated with intentional self-interest, which increases the perceived appropriateness of rejection. Since low offers constitute a clear violation of fairness benchmarks, they may be interpreted as intentionally self-serving when implemented by an LLM. In contrast, high offers do not violate fairness norms. As a result, the effect of an LLM Proposer on the perceived appropriateness of rejection is expected to be stronger for low offers.

**Hypothesis 7:** The upward shift in appropriateness of rejection in the LLM–Proposer/Human–Responder condition (relative to the Human–Proposer/Human–Responder condition) is stronger in "low" splits (when 0-4 is offered to the Responder) than "high" splits (when 6-10 is offered to the Responder).

Together, Hypotheses 6 and 7 complement Hypotheses 4 and 5 by testing whether and how normative standards for rejection differ depending on whether an LLM occupies the role of Proposer or Responder, and whether these differences are more pronounced in situations involving fairness violations.

## Methodology

### *Experimental design*

The experiment followed a similar design to that in Experiment 1, with a few minor differences. Firstly, we used the Ultimatum Game instead of the Dictator Game, and not from the perspective of the first player (the Proposer), but of the second player (the Responder). Secondly, there were three between-subject conditions instead of two. Participants were randomly allocated to one of three conditions, in which: (1) the Proposer was a human and the Responder was a human, (2) the Proposer has delegated the decision to an LLM and the Responder was a human, (3) the Proposer was a human and the Responder was an LLM.

### *Participants*

We recruited a representative sample of people from the United States through Prolific (using the “Representative sample” option on that platform, which ensured that the sample was representative of the US population on sex, age, and ethnicity based on 2021 census data). We aimed to recruit 516 people in each of the three between-subjects conditions, resulting in a sample of 1548 people. The sample size was based on a power simulation, which was intended to obtain 80% statistical power at $\alpha = 5\%$ to detect a Cohen's $d = 0.30$ effect in the between-subjects condition. We amended the parameters used in the simulation due to resource constraints (Lakens, 2022). Because Experiment 2 required three conditions rather than two, we targeted a slightly larger minimum detectable effect ($d = 0.30$) to keep the total sample size within our budget; this still allows detection of small-to-medium effects (Cohen, 1988), which we considered the smallest effects of substantive interest. Given that there are 11 possible splits in the Ultimatum Game and thus 11 tests, we used the Holm-Bonferroni method to counter increased Type I error rates. Our analysis showed that 490 participants per condition would ensure >80% power after the Holm-Bonferroni correction; the code used to perform the power analysis is available at OSF. We recruited more participants per condition to account for a ~5% exclusion rate of inattentive participants. We opted to use just one and not two attention checks to exclude inattentive participants. Ultimately, we excluded 43 participants (2.8% of the full sample) who did not pass the attention check, leaving a final sample of 1505 participants ($M_{age} = 45.9$, $SD = 15.8$; 50.0% female, 48.4% male, 1.6% non-binary or undisclosed).

Participants were paid a flat fee equivalent to £0.60 and told that they could earn an additional amount equivalent to £0.40 if they successfully identified the most common answer in a randomly selected choice from the sample. All sessions took place in October 2025.

### *Statistical analysis*

The statistical analysis largely mirrored that in Experiment 1, with some crucial differences. First, there were three between-subjects conditions (*Human Proposer-Human Responder*, *LLM Proposer-Human Responder*, *Human Proposer-LLM Responder*) instead of two. Second, we did not formally test whether the most socially appropriate splits were the same in all conditions, given that past research suggests that there is no clear social norm for how the Responder should behave (Kimbrough & Vostroknutov, 2016).

### *Open science practices*

The experiment was pre-registered on AsPredicted at https://aspredicted.org/9xt54u.pdf. Data, code, and materials are available at https://osf.io/2yt9e/overview?view_only=8268c7ac95d94f2390fd3ed6dc76269a.

## Results

### *Human is Proposer and LLM is Responder*

First, we present results of a comparison of the Human–LLM condition with the Human–Human condition. The results of a clustered Wilcoxon rank sum test suggest that considering appropriateness ratings in all of the splits, ratings for human and LLM Responders were not significantly different ($z = –0.72$, $p = .472$). As earlier, we complement this omnibus test with a series of Wilcoxon rank sum tests for each of the 11 splits (see **Table 3**).

Using Cliff's Δ there were no substantial differences in the effect in "low" splits ($\Delta_{low\ (0–4)} = –0.043$) and "high" splits ($\Delta_{high\ (6–10)} = 0.016$). Given that the 95% CI for the participant–clustered bootstrapping test contained zero ($\Delta_{high–low} = 0.059$ [–0.025, 0.143]), we conclude that the differences are not statistically significant.

### *LLM is Proposer and Human is Responder*

Next, we compared differences in social appropriateness ratings in the LLM–Human and Human–Human pairs. The results of a clustered Wilcoxon rank sum test suggest that considering appropriateness ratings in all of the splits, ratings for human and LLM Proposers were significantly different ($z = 2.56$, $p = 0.011$).

Using Cliff's Δ there were substantial differences in the effect in "low" splits ($\Delta_{low\ (0–4)} = –0.012$) and "high" splits ($\Delta_{high\ (6–10)} = 0.109$). Given that the 95% CI for the participant-clustered bootstrapping test did not contain zero ($\Delta_{high–low} = 0.121$ [0.037, 0.204]), we conclude that the differences are statistically significant.

**Table 3**

*Difference in social appropriateness ratings – Experiment 2*

| **Amount offered to Responder** | **Human Proposer–LLM Responder vs Control (Human Proposer–Human Responder)** | | | | **LLM Proposer–Human Responder vs Control (Human Proposer–Human Responder)** | | | |
|---|---|---|---|---|---|---|---|---|
| | *Difference in means* | *95% CI* | *W* | *p* | *Difference in means* | *95% CI* | *W* | *p* |
| 0 | –0.106 | [–0.200, –0.011] | 115444 | .546 | –0.086 | [–0.177, 0.005] | 119975 | .111 |
| 1 | –0.081 | [–0.169, 0.006] | 114243 | >.99 | –0.072 | [–0.157, 0.013] | 118251 | .339 |
| 2 | –0.088 | [–0.170, –0.005] | 115771 | .576 | –0.036 | [–0.117, 0.045] | 114386 | .932 |
| 3 | –0.060 | [–0.136, 0.016] | 113745 | >.99 | 0.013 | [–0.063, 0.090] | 109348 | .932 |
| 4 | 0.015 | [–0.057, 0.087] | 105844 | >.99 | 0.082† | [0.009, 0.154] | 101729 | .093 |
| 5 | 0.055 | [–0.037, 0.146] | 104139 | >.99 | 0.064 | [–0.025, 0.153] | 104367 | .339 |
| 6 | 0.047 | [–0.034, 0.128] | 104093 | >.99 | **0.118*** | [0.036, 0.199] | 99738 | .026 |
| 7 | 0.009 | [–0.071, 0.089] | 106360 | >.99 | **0.119*** | [0.037, 0.202] | 99323 | .020 |
| 8 | 0.019 | [–0.063, 0.100] | 105688 | >.99 | **0.150**** | [0.065, 0.236] | 97675 | .004 |
| 9 | 0.004 | [–0.078, 0.086] | 106455 | >.99 | **0.168**** | [0.079, 0.257] | 97662 | .003 |
| 10 | –0.008 | [–0.094, 0.078] | 109323 | >.99 | **0.169**** | [0.077, 0.261] | 97860 | .003 |

*Note.* For each split – separately for the Human-LLM and LLM-Human pairs – we performed a Mann-Whitney-Wilcoxon test. Human-Human is the control condition, and thus negative values indicate that social appropriateness ratings were lower in the Human-LLM/LLM-Human pair. Given that there were 11 splits and for each of them we performed a separate test, so we applied a Holm-Bonferroni correction of the *p*-values. Bold differences are statistically significant at the 5% level or lower.

† $p < .10$, * $p < .05$, ** $p < .01$, *** $p < .001$.

**Discussion**

In Experiment 2, we examined whether injunctive evaluations of rejection decisions in the Ultimatum Game depend on whether the relevant action is implemented by a human or by an LLM acting on behalf of a human. We find no reliable evidence that rejection decisions are evaluated differently when the Responder is an LLM rather than a human, inconsistent with Hypothesis 4. The omnibus comparison between the Human-Proposer/LLM-Responder and Human-Proposer/Human-Responder conditions was not significant, and the stronger effect for low offers predicted in Hypothesis 6 did not emerge. Within the limits of our design, this suggests that replacing a human Responder with an LLM does not substantially alter social appropriateness of rejection decisions. While we are unable to uncover the mechanism behind this pattern, it is worth engaging in some speculation. From a mind perception lens (Gray et al., 2007), we interpret this as consistent with LLMs possessing sufficient agency to enforce norms on behalf of humans. It is less clear how this relates to the experience dimension of mind perception. On the one hand, it could either point to experience of LLMs being high enough to legitimately carry out the role of an enforcer of norms. On the other hand, LLMs may still be perceived as low in experience, but this limitation may be overlooked when the decision is made explicitly on behalf of a human, who possesses such experience.

By contrast, the identity of the Proposer does matter. Rejecting offers made by an LLM Proposer was judged as more appropriate overall than rejecting identical offers made by a human Proposer, supporting Hypothesis 5. Importantly, however, this difference is concentrated among relatively generous offers, not among the low offers for which we had predicted the strongest effect (Hypothesis 7). This pattern is consistent with several *post hoc* interpretations, but our data do not distinguish among them. Observers may treat generous offers from an LLM as misaligned with the interests of the human principal, as strategically odd, or as less legitimate than the same offer made directly by a human. Generous offers in the Ultimatum Game are unexpected and make people question the intentions of the person behind them (Hennig-Schmidt et al., 2008), perhaps even more so when carried out via a machine. Because we did not measure inferred motives, perceived agency, or role legitimacy, it is unclear which of these – of which combination of these – is responsible for the observed pattern. However, in the General Discussion we provide an integrative interpretation of our findings.

## General Discussion

Across two pre-registered, incentivized, and well-powered experimental studies ($N$ = 2,658), we investigate whether people apply the same injunctive norm profiles to humans and to large language models (LLMs) that act on behalf of humans. We use a norm-elicitation procedure (Krupka & Weber, 2013) in which participants rate the social appropriateness of resource allocations. We use two canonical economic games – the Dictator Game and the Ultimatum Game (Güth et al., 1982; Kahneman et al., 1986) – as stylized settings of social interaction that capture different psychological processes (Thielmann et al., 2021).

In our paper, LLM behavior is assessed across allocation and response roles: deciding how to split resources in the Dictator Game, making offers in the Ultimatum Game, and deciding whether to reject offers in the Ultimatum Game. Our results show that although the equal split remains the peak of the norm profile in the Dictator Game, selfish allocations implemented by LLMs are evaluated more strictly than identical allocations made by humans. In the Ultimatum Game, by contrast, rejecting offers from an LLM Proposer is perceived as more socially appropriate than rejecting identical offers made by a human, particularly for relatively generous splits. However, the social appropriateness of rejections by an LLM Responder acting on behalf of a human does not differ from identical rejections by a human Responder.

### A *post hoc* bounded agency interpretation

Overall, we find that people in specific cases apply stricter normative evaluations to LLMs acting on behalf of humans than to humans themselves. As a *post hoc* synthesis, one interpretation of our findings is that evaluations of LLM behavior may depend on the role it occupies and on what observers may infer about the human principal behind the decision when it is delegated.

In allocation settings, self-serving outcomes implemented by LLMs may generate stronger negative evaluations, potentially because of possible suspicion about the motives of the human behind it. Machines have the potential to be intentionally used by humans to offset blame for making an unfavorable offer (Cohn et al., 2022; Köbis et al., 2021). Observers may view the LLM as a less independent beneficiary and more as an instrument acting on behalf of the interests of the human. This interpretation is consistent with our finding that selfish allocations in the Dictator Game were evaluated more harshly when implemented by an LLM than by a human, even though the equal split remained the most appropriate response.

At the same time, our results from the Ultimatum Game suggest that evaluations are role-dependent rather than universal. Rejection decisions implemented by an LLM Responder in the Ultimatum Game were judged no less appropriate than identical rejection decisions implemented by a human Responder. This suggests that some actions – potentially related to enforcement – are seen as normatively acceptable when carried out by an LLM acting on behalf of a human.

We consider our findings broadly consistent with the idea that LLMs may be treated as having a *bounded form of agency*, whose normative acceptability depends on the role they occupy and on what observers infer about the principal behind them. However, since we did not directly measure inferred motives, perceived agency, or role legitimacy, this interpretation should be treated as a theoretically grounded synthesis rather than as a tested mechanism.

We describe the *bounded agency* interpretation in more detail in the **Supplementary Materials**.

### Reconciliation with prior research

We compare our findings with prior work where machines occupied either the Proposer or the Responder role in the Ultimatum Game. In the work by Sanfey et al. (2003), participants reacted more adversely to low offers from humans than from computers. The authors attributed this to the fact that computers lack agency and intent. More recent work by Borthakur et al. (2025) find that low offers are more likely to be rejected when they come from an AI agent than from a human, and high offers are less likely to be rejected when they come from an AI agent than from a human. By contrast, our results show no difference in the perceived appropriateness of rejecting low offers, and we observe the opposite pattern for high offers, where participants evaluate the rejection of relatively generous offers as more appropriate when these offers were implemented by an LLM.

It is well-known that while people may know what the appropriate way is to behave in a given situation – what *ought* to be done – it does not mean that they will follow it (Blake, 2018; Gächter et al., 2017). Prior studies examine actual behavior, whereas we investigate injunctive social norms governing such behavior. Our results relate to evaluative standards rather than behavioral responses, and do not imply that participants would make rejection decisions themselves. Whether differences in norm evaluations translate into differences in behavior, delegation preferences, or acceptance of decisions remains an open empirical question.

Another important distinction with previous studies is that we explicitly told participants in our experiment that the machine is acting on behalf of a human, whereas in Sanfey et al. (2003) and Borthakur et al. (2025) participants interact only with a machine, without any mention that there is a human as the final beneficiary. Prior research has shown that it is important to explicitly mention human beneficiaries behind the machine if they are present, instead of expecting that people will infer it (von Schenk et al., 2023).

Finally, our study focuses on a one-shot interaction, whereas some earlier studies consider repeated settings, which may also contribute to differences in findings.

**Practical implications**

Along with the proliferation of easily-accessible AI systems such as LLMs, the social stakes of machine behaviors are increasing (Hadfield & Koh, 2025; Rahwan et al., 2019). While economic games are simplified environments, they capture real-life interactions, allowing us to draw some practical implications for our findings.

One type of behavior resembling the Dictator Game is an employer at work deciding how much bonus to pay employees based on their performance. These findings suggest that, in analogous settings such as bonus allocation, unfavorable outcomes implemented by AI systems may be seen as less appropriate. For example, employees receiving a lower bonus than expected – which could otherwise have been viewed as fair – may interpret AI-mediated implementation more negatively, potentially because it is seen as strategically motivated or impersonal. It is important to note that any implications for outcomes such as job satisfaction remain to be tested, as our results concern normative evaluations rather than behavior or downstream consequences.

Beyond resource allocation, machines are delegated roles that involve enforcing rules, deciding, for instance, whether economic transactions or claims should be accepted or rejected, analogously to the Responder's role in the Ultimatum Game. For example, algorithms are used to evaluate welfare eligibility (Dong et al., 2025) and insurance claims (Rucker et al., 2023). Our findings suggest that, once a decision to delegate *has been made*, the perceived appropriateness of rejection does not differ depending on whether the rejection is implemented by a human or a machine acting on its behalf. One possible interpretation is that the role of the Responder does not trigger the same inferences as the allocator role, such that delegation does not alter normative evaluations of rejection in the same way it alters the evaluation of allocations.

***Implications for AI design***

Our findings can complement technical and governance work on AI by identifying cases in which decisions by AI agents may face greater normative scrutiny than equivalent decisions made by humans. Elicited injunctive norms can serve as an input in discussions about how AI systems should behave in settings that have social consequences.

Our results indicate that identical allocation decisions are evaluated differently depending on whether they are implemented by a human or by a machine acting on behalf of a human. In particular, allocations that deviate from the commonly accepted fairness benchmarks (e.g., equal split) are judged more harshly when carried out by a machine. This suggests that AI-mediated decisions may be subject to stricter normative standards, especially when the outcome appears self-serving.

Furthermore, our results from the Ultimatum Game indicate that offers above the fairness benchmark may also be subject to a stricter standard evaluation when implemented by a machine, as participants judged rejection of such offers to be more appropriate. Therefore, our results point towards a narrower range of behavior that is perceived as appropriate when decisions are made by AI systems acting on behalf of humans. This might be especially important for people with a high individual propensity to adhere to norms, for whom a misbehaving machine (see, e.g., Basu, 2026) will be especially problematic.

## Limitations

Our study has several limitations. First, we elicit only injunctive norms and not actual behavior. We therefore cannot infer whether these judgments translate into allocation choices, rejection behavior, willingness to delegate to AI, or support for AI use in these roles. Second, we did not directly measure inferred motives, perceived agency, or role legitimacy. Thus, for example, while we interpret our findings as consistent with the notion that people infer the motives behind the delegation, we cannot determine whether inferred motives cause the stricter normative evaluations of LLMs for selfish splits in the Dictator Game. Third, in Experiment 2, we elicited appropriateness ratings for rejection decisions for each offer, and not acceptance decisions: if the rejection of a particular offer is socially appropriate, it does not immediately mean that the acceptance of the offer will be socially appropriate. Thus, it is unclear if key findings would be mirrored when eliciting the appropriateness of acceptance.

Additionally, although we use two economic games, they still represent only a subset of canonical economic games (Thielmann et al., 2021). Future research could investigate norms concerning behavior in other games, such as the Trust Game or the Public Goods Game. Past research has uncovered what behavior is considered appropriate in such games (Kimbrough & Vostroknutov, 2016), which provides a solid basis for deriving expectations about how humans should behave in situations modeled by these games. Such investigations would be valuable, as they could test whether the patterns observed here generalize to settings in which AI systems make primary allocation decisions in comparison to settings in which they evaluate or respond to others' decisions.

A more general concern regarding the use of economic games is their external validity (Levitt & List, 2007). Economic games are useful abstractions that distill complex behaviors into simpler ones that can be clearly and quickly understood by participants. However, there are legitimate questions about how well they map onto real-world behaviors. Moreover, these games are sensitive to design details and possible outcomes. For example, people behave less prosocially in a Dictator Game after the possibility of taking away (and not only giving) is introduced into the game (List, 2007).

Finally, we studied only people from Western cultures and highly developed countries, and our samples – while representative of sex, age and ethnicity – may not have been representative on other factors such as education or experience with AI, due to participants self-selecting into online experiments (Pauketat et al., 2026; Tang et al., 2022). Replication in non-Western countries is therefore important before drawing broad generalizations. It is well known that attitudes towards AI differ between countries (Gillespie et al., 2025; Zhang, 2024), which further highlights the need for cross-cultural validation of our findings.

## Future research directions

To conserve space, we present directions for future research in the **Supplementary Materials**.

# Conclusions

Large language models are increasingly used to assist or implement decisions that have social consequences (Hadfield & Koh, 2025). Using two economic games – the Dictator Game and the Ultimatum Game – we show that people sometimes apply different social norms to humans and to LLMs acting on their behalf. We interpret our findings as consistent with *bounded* agency, in which evaluations depend on whether the actions of the LLM align with the presumed intentions of the human behind it and with the role the system occupies. As AI systems are increasingly used to allocate resources and enforce the fairness of allocations and rules on behalf of individuals or organizations (Hadfield & Koh, 2025), the knowledge about when people apply stricter (or the same) normative standards for LLMs can inform the design of AI systems. This work complements the literature on machine behavior (Rahwan et al., 2019).

# Supplementary Materials

## Do people apply different norms to humans and large language models acting on their behalf? Evidence from norm elicitations in two canonical economic games

## Table of contents

## Part A. Supplementary tables and figures

**Table S1**

*Descriptive statistics*

| | **Experiment 1 (*N* = 1160, UK)** | **Experiment 2 (*N* = 1548, US)** |
|---|---|---|
| | ***M (SD) or n (%)*** | ***M (SD) or n (%)*** |
| **Age** | 47.02 (15.09) | 45.80 (15.77) |
| **Gender** | | |
| Female | 582 (50.2%) | 767 (49.5%) |
| Male | 566 (48.8%) | 757 (48.9%) |
| Non-binary | 7 (0.6%) | 19 (1.2%) |
| Prefer not to say | 5 (0.4%) | 5 (0.3%) |
| **Education** | | |
| Primary education | 5 (0.4%) | 71 (4.6%) |
| Secondary education | 430 (37.1%) | 530 (34.2%) |
| Bachelor (or equivalent) | 484 (41.7%) | 657 (42.4%) |
| Masters (or equivalent) | 205 (17.7%) | 229 (14.8%) |
| Doctoral (or equivalent) | 36 (3.1%) | 61 (3.9%) |
| **Occupation** | | |
| Employed - Private Sector | 449 (38.7%) | 679 (43.9%) |
| Employed - Public Sector | 239 (20.6%) | 247 (16.0%) |
| Self-employed | 150 (12.9%) | 203 (13.1%) |
| Student | 34 (2.9%) | 41 (2.6%) |
| Not currently employed | 288 (24.8%) | 378 (24.4%) |
| **What participant has studied** | | |
| Economics / Business Administration / Finance / Accounting / Marketing | 209 (18.0%) | 362 (23.4%) |
| I do not have a degree nor I am currently pursuing one | 346 (29.8%) | 382 (24.7%) |
| Other (Non-Economics related studies) | 605 (52.2%) | 804 (51.9%) |
| **Income** | | |
| I do not have an income | 9 (0.8%) | 15 (1.0%) |
| Less than 20,000 | 308 (26.6%) | 271 (17.5%) |
| Between 20,000-59,999 | 664 (57.2%) | 585 (37.8%) |
| Between 60,000-99,999 | 112 (9.7%) | 384 (24.8%) |
| More than 100,000 | 18 (1.6%) | 254 (16.4%) |
| Prefer not to disclose | 49 (4.2%) | 39 (2.5%) |
| **Trust in AI** | | |
| 1 = I do not trust LLMs at all | 105 (9.1%) | 167 (10.8%) |
| 2 = I trust LLMs less than humans | 628 (54.1%) | 701 (45.3%) |
| 3 = I trust LLMs the same as humans | 323 (27.8%) | 473 (30.6%) |
| 4 = I trust LLMs more than humans | 98 (8.4%) | 175 (11.3%) |
| 5 = I fully trust LLMs | 6 (0.5%) | 32 (2.1%) |
| **Use of AI** | | |
| Never | 152 (13.1%) | 140 (9.0%) |
| Rarely | 368 (31.7%) | 382 (24.7%) |
| A few days per week | 435 (37.5%) | 634 (41.0%) |
| Every day | 205 (17.7%) | 392 (25.3%) |
| **Experience with game** | 51 (4.4%) | 170 (11.0%) |

*Note.* Income in Experiments 1 and 2 was in pounds and dollars, respectively.

**Table S2**

*Linear mixed model regressions*

| | Dictator Game | | | Ultimatum Game: Human-LLM | | | Ultimatum Game: LLM-Human | | |
|---|---|---|---|---|---|---|---|---|---|
| *Predictors* | *Estimates* | *95% CI* | *p* | *Estimates* | *95% CI* | *p* | *Estimates* | *95% CI* | *p* |
| 0 given/offered (baseline) | **-0.75*** ** | [-0.80, -0.70] | <0.001 | **0.48*** ** | [0.42, 0.54] | <0.001 | **0.48*** ** | [0.42, 0.54] | <0.001 |
| 1 given/offered | **0.13*** ** | [0.07, 0.19] | <0.001 | **-0.07* ** | [-0.15, -0.00] | 0.041 | **-0.07* ** | [-0.15, -0.00] | 0.041 |
| 2 given/offered | **0.33*** ** | [0.27, 0.39] | <0.001 | **-0.16*** ** | [-0.24, -0.09] | <0.001 | **-0.16*** ** | [-0.24, -0.09] | <0.001 |
| 3 given/offered | **0.60*** ** | [0.54, 0.66] | <0.001 | **-0.31*** ** | [-0.38, -0.24] | <0.001 | **-0.31*** ** | [-0.38, -0.24] | <0.001 |
| 4 given/offered | **0.93*** ** | [0.87, 0.99] | <0.001 | **-0.50*** ** | [-0.58, -0.43] | <0.001 | **-0.50*** ** | [-0.58, -0.43] | <0.001 |
| 5 given/offered | **1.53*** ** | [1.48, 1.59] | <0.001 | **-0.84*** ** | [-0.91, -0.77] | <0.001 | **-0.84*** ** | [-0.91, -0.77] | <0.001 |
| 6 given/offered | **1.27*** ** | [1.21, 1.33] | <0.001 | **-0.95*** ** | [-1.02, -0.88] | <0.001 | **-0.95*** ** | [-1.02, -0.88] | <0.001 |
| 7 given/offered | **1.12*** ** | [1.06, 1.18] | <0.001 | **-0.99*** ** | [-1.06, -0.92] | <0.001 | **-0.99*** ** | [-1.06, -0.92] | <0.001 |
| 8 given/offered | **0.99*** ** | [0.93, 1.05] | <0.001 | **-1.04*** ** | [-1.11, -0.97] | <0.001 | **-1.04*** ** | [-1.11, -0.97] | <0.001 |
| 9 given/offered | **0.92*** ** | [0.86, 0.97] | <0.001 | **-1.07*** ** | [-1.14, -1.00] | <0.001 | **-1.07*** ** | [-1.14, -1.00] | <0.001 |
| 10 given/offered | **0.88*** ** | [0.82, 0.94] | <0.001 | **-1.06*** ** | [-1.13, -0.99] | <0.001 | **-1.06*** ** | [-1.13, -0.99] | <0.001 |
| 0 x LLM-involvement (baseline for interaction) | **-0.10** ** | [-0.17, -0.03] | 0.006 | **-0.11* ** | [-0.19, -0.02] | 0.013 | **-0.09* ** | [-0.17, -0.00] | 0.045 |
| 1 x LLM-involvement | -0.03 | [-0.12, 0.05] | 0.463 | 0.02 | [-0.08, 0.13] | 0.638 | 0.01 | [-0.09, 0.12] | 0.787 |
| 2 x LLM-involvement | -0.07 | [-0.15, 0.01] | 0.097 | 0.02 | [-0.08, 0.12] | 0.725 | 0.05 | [-0.05, 0.15] | 0.327 |
| 3 x LLM-involvement | -0.08 | [-0.17, 0.00] | 0.056 | 0.05 | [-0.06, 0.15] | 0.378 | 0.10 | [-0.00, 0.20] | 0.053 |
| 4 x LLM-involvement | **-0.11* ** | [-0.19, -0.03] | 0.011 | **0.12* ** | [0.02, 0.22] | 0.019 | **0.17** ** | [0.07, 0.27] | 0.001 |

| | Dictator Game | | | Ultimatum Game: Human-LLM | | | Ultimatum Game: LLM-Human | | |
|---|---|---|---|---|---|---|---|---|---|
| *Predictors* | *Estimates* | *95% CI* | *p* | *Estimates* | *95% CI* | *p* | *Estimates* | *95% CI* | *p* |
| 5 x LLM-involvement | **-0.10*** | [-0.18, -0.01] | 0.025 | **0.16**** | [0.06, 0.26] | 0.002 | **0.15**** | [0.05, 0.25] | 0.004 |
| 6 x LLM-involvement | -0.02 | [-0.11, 0.06] | 0.573 | **0.15**** | [0.05, 0.25] | 0.003 | **0.20***** | [0.10, 0.30] | <0.001 |
| 7 x LLM-involvement | -0.00 | [-0.08, 0.08] | 0.985 | **0.11*** | [0.01, 0.22] | 0.027 | **0.21***** | [0.10, 0.31] | <0.001 |
| 8 x LLM-involvement | 0.04 | [-0.05, 0.12] | 0.395 | **0.12*** | [0.02, 0.23] | 0.016 | **0.24***** | [0.14, 0.34] | <0.001 |
| 9 x LLM-involvement | 0.06 | [-0.03, 0.14] | 0.199 | **0.11*** | [0.01, 0.21] | 0.034 | **0.25***** | [0.15, 0.36] | <0.001 |
| 10 x LLM-involvement | **0.17***** | [0.09, 0.26] | <0.001 | 0.10 | [-0.00, 0.20] | 0.059 | **0.26***** | [0.15, 0.36] | <0.001 |
| $\sigma^2$ | 0.26 | | | 0.34 | | | 0.33 | | |
| $\tau$00 [participant_id] | 0.12 | | | 0.12 | | | 0.13 | | |
| ICC | 0.30 | | | 0.26 | | | 0.28 | | |
| N [participant_id] | 1153 | | | 1001 | | | 1003 | | |
| Observations | 12683 | | | 11011 | | | 11033 | | |
| Marginal $R^2$ | 0.36 | | | 0.25 | | | 0.23 | | |
| Conditional $R^2$ | 0.56 | | | 0.44 | | | 0.45 | | |

*Note.* Estimates where obtained using the *lme4* (Bates et al., 2015) packages in *R*. Given that there are 11 splits and for each of them we perform a separate test, we performed a Holm-Bonferroni correction of the *p*-values. Bold estimates are statistically significant at the 5% level or lower.
† p < .10, * p < .05, ** p < .01, *** p < .001

**Table S3**

*Linear mixed model regressions with controls for age, gender, education, occupation, studies, income, trust in AI, use of AI, experience of game (Dictator or Ultimatum Game)*

| | Dictator Game | | | Ultimatum Game: Human-LLM | | | Ultimatum Game: LLM-Human | | |
|---|---|---|---|---|---|---|---|---|---|
| *Predictors* | *Estimates* | *95% CI* | *p* | *Estimates* | *95% CI* | *p* | *Estimates* | *95% CI* | *p* |
| 0 given/offered (baseline) | **-0.74*** ** | [-0.87, -0.61] | <0.001 | **0.39*** ** | [0.23, 0.54] | <0.001 | **0.47*** ** | [0.31, 0.62] | <0.001 |
| 1 given/offered | **0.13*** ** | [0.07, 0.19] | <0.001 | **-0.07* ** | [-0.15, -0.00] | 0.041 | **-0.07* ** | [-0.15, -0.00] | 0.041 |
| 2 given/offered | **0.33*** ** | [0.27, 0.39] | <0.001 | **-0.16*** ** | [-0.24, -0.09] | <0.001 | **-0.16*** ** | [-0.24, -0.09] | <0.001 |
| 3 given/offered | **0.60*** ** | [0.54, 0.66] | <0.001 | **-0.31*** ** | [-0.38, -0.24] | <0.001 | **-0.31*** ** | [-0.38, -0.24] | <0.001 |
| 4 given/offered | **0.93*** ** | [0.87, 0.99] | <0.001 | **-0.50*** ** | [-0.58, -0.43] | <0.001 | **-0.50*** ** | [-0.58, -0.43] | <0.001 |
| 5 given/offered | **1.53*** ** | [1.48, 1.59] | <0.001 | **-0.84*** ** | [-0.91, -0.77] | <0.001 | **-0.84*** ** | [-0.91, -0.77] | <0.001 |
| 6 given/offered | **1.27*** ** | [1.21, 1.33] | <0.001 | **-0.95*** ** | [-1.02, -0.88] | <0.001 | **-0.95*** ** | [-1.02, -0.88] | <0.001 |
| 7 given/offered | **1.12*** ** | [1.06, 1.18] | <0.001 | **-0.99*** ** | [-1.06, -0.92] | <0.001 | **-0.99*** ** | [-1.06, -0.92] | <0.001 |
| 8 given/offered | **0.99*** ** | [0.93, 1.05] | <0.001 | **-1.04*** ** | [-1.11, -0.97] | <0.001 | **-1.04*** ** | [-1.11, -0.97] | <0.001 |
| 9 given/offered | **0.92*** ** | [0.86, 0.97] | <0.001 | **-1.07*** ** | [-1.14, -1.00] | <0.001 | **-1.07*** ** | [-1.14, -1.00] | <0.001 |
| 10 given/offered | **0.88*** ** | [0.82, 0.94] | <0.001 | **-1.06*** ** | [-1.13, -0.99] | <0.001 | **-1.06*** ** | [-1.13, -0.99] | <0.001 |
| 0 x LLM-involvement (baseline for interaction) | **-0.10** ** | [-0.17, -0.03] | 0.006 | **-0.10* ** | [-0.18, -0.01] | 0.022 | **-0.09* ** | [-0.17, -0.00] | 0.043 |
| 1 x LLM-involvement | -0.03 | [-0.12, 0.05] | 0.463 | 0.02 | [-0.08, 0.13] | 0.638 | 0.01 | [-0.09, 0.12] | 0.787 |
| 2 x LLM-involvement | -0.07 | [-0.15, 0.01] | 0.097 | 0.02 | [-0.08, 0.12] | 0.725 | 0.05 | [-0.05, 0.15] | 0.327 |

| | Dictator Game | | | Ultimatum Game: Human-LLM | | | Ultimatum Game: LLM-Human | | |
|---|---|---|---|---|---|---|---|---|---|
| *Predictors* | *Estimates* | *95% CI* | *p* | *Estimates* | *95% CI* | *p* | *Estimates* | *95% CI* | *p* |
| 3 x LLM-involvement | -0.08 | [-0.17, 0.00] | 0.056 | 0.05 | [-0.06, 0.15] | 0.378 | 0.10 | [-0.00, 0.20] | 0.053 |
| 4 x LLM-involvement | **-0.11*** | [-0.19, -0.03] | 0.011 | **0.12*** | [0.02, 0.22] | 0.019 | **0.17**** | [0.07, 0.27] | 0.001 |
| 5 x LLM-involvement | **-0.10*** | [-0.18, -0.01] | 0.025 | **0.16**** | [0.06, 0.26] | 0.002 | **0.15**** | [0.05, 0.25] | 0.004 |
| 6 x LLM-involvement | -0.02 | [-0.11, 0.06] | 0.573 | **0.15**** | [0.05, 0.25] | 0.003 | **0.20***** | [0.10, 0.30] | <0.001 |
| 7 x LLM-involvement | -0.00 | [-0.08, 0.08] | 0.985 | **0.11*** | [0.01, 0.22] | 0.027 | **0.21***** | [0.10, 0.31] | <0.001 |
| 8 x LLM-involvement | 0.04 | [-0.05, 0.12] | 0.395 | **0.12*** | [0.02, 0.23] | 0.016 | **0.24***** | [0.14, 0.34] | <0.001 |
| 9 x LLM-involvement | 0.06 | [-0.03, 0.14] | 0.199 | **0.11*** | [0.01, 0.21] | 0.034 | **0.25***** | [0.15, 0.36] | <0.001 |
| 10 x LLM-involvement | **0.17***** | [0.09, 0.26] | <0.001 | 0.10 | [-0.00, 0.20] | 0.059 | **0.26***** | [0.15, 0.36] | <0.001 |
| ***CONTROLS*** | | | | | | | | | |
| Age | -0.00 | [-0.00, 0.00] | 0.654 | 0.00 | [-0.00, 0.00] | 0.635 | -0.00 | [-0.00, 0.00] | 0.821 |
| Gender [Male] | 0.03 | [-0.02, 0.07] | 0.271 | 0.02 | [-0.03, 0.07] | 0.520 | 0.03 | [-0.02, 0.08] | 0.263 |
| Gender [Non-binary] | 0.03 | [-0.25, 0.31] | 0.843 | -0.17 | [-0.41, 0.06] | 0.143 | -0.13 | [-0.35, 0.08] | 0.227 |
| Gender [Prefer not to say] | 0.12 | [-0.22, 0.45] | 0.492 | 0.12 | [-0.28, 0.52] | 0.552 | 0.38 | [-0.02, 0.79] | 0.063 |
| Education [Doctoral (or equivalent)] | 0.03 | [-0.10, 0.16] | 0.623 | -0.03 | [-0.16, 0.09] | 0.614 | -0.06 | [-0.20, 0.07] | 0.360 |
| Education [Masters (or equivalent)] | 0.02 | [-0.05, 0.08] | 0.635 | 0.03 | [-0.05, 0.10] | 0.463 | -0.01 | [-0.09, 0.07] | 0.834 |
| Education [Primary education] | -0.00 | [-0.38, 0.37] | 0.982 | -0.03 | [-0.16, 0.11] | 0.694 | 0.06 | [-0.08, 0.20] | 0.405 |
| Education [Secondary education] | 0.00 | [-0.08, 0.08] | 0.996 | 0.01 | [-0.06, 0.08] | 0.828 | -0.01 | [-0.08, 0.06] | 0.786 |
| Occupation [Employed - Public Sector] | -0.02 | [-0.09, 0.04] | 0.435 | 0.05 | [-0.02, 0.12] | 0.199 | 0.05 | [-0.02, 0.12] | 0.178 |

| | **Dictator Game** | | | **Ultimatum Game: Human-LLM** | | | **Ultimatum Game: LLM-Human** | | |
|---|---|---|---|---|---|---|---|---|---|
| *Predictors* | *Estimates* | *95% CI* | *p* | *Estimates* | *95% CI* | *p* | *Estimates* | *95% CI* | *p* |
| Occupation [Not currently employed] | 0.04 | [-0.03, 0.10] | 0.251 | 0.02 | [-0.06, 0.09] | 0.667 | 0.03 | [-0.04, 0.11] | 0.367 |
| Occupation [Self-employed] | 0.05 | [-0.02, 0.13] | 0.176 | 0.01 | [-0.07, 0.08] | 0.891 | 0.01 | [-0.07, 0.10] | 0.737 |
| Occupation [Student] | 0.06 | [-0.08, 0.20] | 0.388 | 0.06 | [-0.09, 0.21] | 0.463 | -0.06 | [-0.23, 0.11] | 0.508 |
| Studies [I do not have a degree nor I am currently pursuing one] | 0.01 | [-0.08, 0.10] | 0.867 | 0.01 | [-0.07, 0.10] | 0.790 | 0.01 | [-0.08, 0.10] | 0.889 |
| Studies [Other (Non-Economics related studies)] | -0.00 | [-0.07, 0.06] | 0.905 | 0.02 | [-0.04, 0.08] | 0.532 | -0.00 | [-0.07, 0.06] | 0.960 |
| Income [Less than 20,000] | -0.01 | [-0.07, 0.05] | 0.745 | 0.04 | [-0.03, 0.12] | 0.272 | **0.08*** | [0.00, 0.16] | 0.043 |
| Income [Between 60,000-99,999] | 0.01 | [-0.07, 0.09] | 0.828 | 0.02 | [-0.04, 0.09] | 0.485 | 0.03 | [-0.04, 0.09] | 0.431 |
| Income [More than 100,000] | 0.12 | [-0.06, 0.29] | 0.200 | 0.01 | [-0.06, 0.09] | 0.710 | 0.02 | [-0.05, 0.10] | 0.561 |
| Income [I do not have an income] | -0.13 | [-0.38, 0.13] | 0.325 | 0.01 | [-0.31, 0.33] | 0.952 | 0.18 | [-0.07, 0.43] | 0.159 |
| Income [Prefer not to disclose] | -0.06 | [-0.17, 0.05] | 0.303 | 0.08 | [-0.09, 0.24] | 0.351 | 0.08 | [-0.08, 0.25] | 0.317 |
| Use of AI [Every day] | -0.06 | [-0.12, 0.01] | 0.089 | -0.04 | [-0.10, 0.03] | 0.259 | -0.04 | [-0.10, 0.03] | 0.245 |
| Use of AI [Never] | -0.05 | [-0.12, 0.02] | 0.190 | 0.05 | [-0.05, 0.14] | 0.331 | 0.02 | [-0.08, 0.11] | 0.748 |
| Use of AI [Rarely] | **-0.06*** | [-0.11, -0.01] | 0.030 | -0.04 | [-0.11, 0.02] | 0.184 | 0.02 | [-0.05, 0.08] | 0.629 |
| Trust in AI | 0.01 | [-0.02, 0.04] | 0.651 | 0.01 | [-0.02, 0.04] | 0.393 | -0.01 | [-0.04, 0.02] | 0.502 |
| Game experience [1 = yes, 0 = no] | -0.03 | [-0.13, 0.08] | 0.643 | 0.00 | [-0.07, 0.08] | 0.910 | 0.03 | [-0.06, 0.11] | 0.530 |
| $\sigma^2$ | 0.26 | | | 0.34 | | | 0.33 | | |
| $\tau_{00}$ [participant_id] | 0.12 | | | 0.12 | | | 0.13 | | |

| | Dictator Game | | | Ultimatum Game: Human-LLM | | | Ultimatum Game: LLM-Human | | |
|---|---|---|---|---|---|---|---|---|---|
| *Predictors* | *Estimates* | *95% CI* | *p* | *Estimates* | *95% CI* | *p* | *Estimates* | *95% CI* | *p* |
| ICC | 0.31 | | | 0.26 | | | 0.28 | | |
| N [participant_id] | 1153 | | | 1001 | | | 1003 | | |
| Observations | 12683 | | | 11011 | | | 11033 | | |
| Marginal $R^2$ | 0.36 | | | 0.25 | | | 0.24 | | |
| Conditional $R^2$ | 0.56 | | | 0.45 | | | 0.45 | | |

*Note.* Estimates where obtained using the *lme4* (Bates et al., 2015) packages in *R*. Given that there are 11 splits and for each of them we perform a separate test, we performed a Holm-Bonferroni correction of the *p*-values. Bold estimates are statistically significant at the 5% level or lower.
† $p < .10$, * $p < .05$, ** $p < .01$, *** $p < .001$

**Table S4**

*Tests of hypotheses using different exclusion criteria*

| | | A. Without controls | | | B. With controls | | |
|---|---|---|---|---|---|---|---|
| ***Experiment*** | ***Hypothesis*** | ***Exclusions based on the amount that was to be split*** | ***Exclusions based on the amount that was to be split and the identity of the Dictator*** | ***Exclusions based on the amount that was to be split and straight-liners*** | ***Exclusions based on the amount that was to be split*** | ***Exclusions based on the amount that was to be split and the identity of the Dictator*** | ***Exclusions based on the amount that was to be split and straight-liners*** |
| Experiment 1 | H1 | Observed peaks: Human = 5, LLM = 5; $\Delta = 0$ 95% CI for $\Delta$: [0, 0] | Observed peaks: Human = 5, LLM = 5; $\Delta = 0$ 95% CI for $\Delta$: [0, 0] | – | Observed peaks: Human = 5, LLM = 5; $\Delta = 0$ 95% CI for $\Delta$: [0, 0] | Observed peaks: Human = 5, LLM = 5; $\Delta = 0$ 95% CI for $\Delta$: [0, 0] | – |
| | H2 | $z = -5.12, p < .001$ | $z = -4.22, p < .001$ | – | $z = -4.20, p = .001$ | $z = -3.27, p = .001$ | – |
| | H3 | $\Delta_{low\ (0–4)} = -0.163$ $\Delta_{high\ (6-10)} = -0.047$ $\Delta_{high-low} = 0.116$ [0.048, 0.183]) | $\Delta_{low\ (0–4)} = -0.161$ $\Delta_{high\ (6-10)} = -0.041$ $\Delta_{high-low} = 0.120$ [0.045, 0.196] | – | $\Delta_{low\ (0–4)} = -0.154$ $\Delta_{high\ (6-10)} = -0.033$ $\Delta_{high-low} = 0.120$ [0.048, 0.191]) | $\Delta_{low\ (0–4)} = -0.143$ $\Delta_{high\ (6-10)} = -0.022$ $\Delta_{high-low} = 0.121$ [0.040, 0.203] | – |
| Experiment 2 | H4 | $z = -0.72, p = .472$ | – | $z = -0.33, p = .744$ | $z = 0.07, p = .947$ | – | $z = 0.94, p = .348$ |
| | H5 | $z = 2.56, p = .011$ | – | $z = 2.43, p = .015$ | $z = 2.68, p = .007$ | – | $z = 2.97, p = .003$ |
| | H6 | $\Delta_{low\ (0–4)} = -0.043$ $\Delta_{high\ (6-10)} = 0.016$ $\Delta_{high-low} = 0.059$ [-0.025, 0.143] | – | $\Delta_{low\ (0–4)} = -0.036$ $\Delta_{high\ (6-10)} = 0.024$ $\Delta_{high-low} = 0.061$ [-0.028, 0.150] | $\Delta_{low\ (0–4)} = -0.025$ $\Delta_{high\ (6-10)} = 0.037$ $\Delta_{high-low} = 0.062$ [-0.025, 0.150] | – | $\Delta_{low\ (0–4)} = -0.010$ $\Delta_{high\ (6-10)} = 0.064$ $\Delta_{high-low} = 0.073$ [-0.021, 0.166] |
| | H7 | $\Delta_{low\ (0–4)} = -0.012$ $\Delta_{high\ (6-10)} = 0.109$ $\Delta_{high-low} = 0.121$ [0.037, 0.204] | – | $\Delta_{low\ (0–4)} = -0.021$ $\Delta_{high\ (6-10)} = 0.109$ $\Delta_{high-low} = 0.129$ [0.043, 0.215] | $\Delta_{low\ (0–4)} = -0.009$ $\Delta_{high\ (6-10)} = 0.124$ $\Delta_{high-low} = 0.13$ [0.046, 0.219] | – | $\Delta_{low\ (0–4)} = -0.008$ $\Delta_{high\ (6-10)} = 0.138$ $\Delta_{high-low} = 0.146$ [0.058, 0.238] |

*Note.* (A) In Experiment 1, we planned to use more strict exclusion criteria: based on the amount that was to be split *and* the identity of the Dictator. However, as the latter might have been unclear to participants – based on a high exclusion rate – in the main body of the text, we report data from the *Exclusions based on the amount that was to be split* column. (B) In this part of the table we present the same analyses that were presented without controls, with one key difference: we corrected ratings for the effect of the control variables. To do so, we first residualized participants' ratings with respect to the available covariates (e.g., demographic characteristics and AI-use/trust measures) by regressing the rating on those controls, taking the residuals, and adding back the grand mean so the adjusted outcome remained on the original scale.

## Table S5

*Estimated marginal effects in linear mixed models after controlling for age, gender, education, occupation, studies, income, trust in AI, use of AI, experience of game (Dictator or Ultimatum Game)– Experiment 1 and Experiment 2*

| Given to Recipient (Experiment 1) or Offered to Responder (Experiment 2) | Dictator Game (Experiment 1): social appropriateness of *making* split by Dictator | | | | Ultimatum Game (Experiment 2): social appropriateness of *rejecting* split by Responder | | | | | | | |
|---|---|---|---|---|---|---|---|---|---|---|---|---|
| | LLM Dictator vs Human Dictator | | | | Human Proposer-LLM Responder vs Human Proposer-Human Responder | | | | LLM Proposer-Human Responder vs Human Proposer-Human Responder | | | |
| | *Estimate* | *95% CI* | *t* | *p* | *Estimate* | *95% CI* | *t* | *p* | *Estimate* | *95% CI* | *t* | *p* |
| 0 | **-0.100**** | [-0.171, -0.028] | -2.725 | 0.006 | **-0.098*** | [-0.182, -0.014] | -2.29 | 0.022 | **-0.088*** | [-0.172, -0.003] | -2.026 | 0.043 |
| 1 | **-0.131***** | [-0.203, -0.059] | -3.584 | <0.001 | -0.074† | [-0.158, 0.010] | -1.721 | 0.085 | -0.074† | [-0.158, 0.011] | -1.703 | 0.089 |
| 2 | **-0.171***** | [-0.242, -0.099] | -4.668 | <0.001 | -0.08† | [-0.164, 0.004] | -1.864 | 0.062 | -0.037 | [-0.122, 0.048] | -0.857 | 0.392 |
| 3 | **-0.182***** | [-0.253, -0.110] | -4.965 | <0.001 | -0.052 | [-0.136, 0.032] | -1.224 | 0.221 | 0.012 | [-0.073, 0.097] | 0.277 | 0.782 |
| 4 | **-0.209***** | [-0.280, -0.137] | -5.707 | <0.001 | 0.023 | [-0.061, 0.107] | 0.541 | 0.589 | 0.080† | [-0.005, 0.165] | 1.851 | 0.064 |
| 5 | **-0.196***** | [-0.268, -0.124] | -5.357 | <0.001 | 0.062 | [-0.022, 0.146] | 1.457 | 0.145 | 0.062 | [-0.023, 0.147] | 1.439 | 0.150 |
| 6 | **-0.124***** | [-0.195, -0.052] | -3.385 | <0.001 | 0.055 | [-0.029, 0.139] | 1.285 | 0.199 | **0.116**** | [0.031, 0.201] | 2.687 | 0.007 |
| 7 | **-0.100**** | [-0.172, -0.029] | -2.746 | 0.006 | 0.017 | [-0.067, 0.101] | 0.391 | 0.696 | **0.118**** | [0.033, 0.203] | 2.726 | 0.006 |
| 8 | -0.063† | [-0.135, 0.008] | -1.729 | 0.084 | 0.026 | [-0.058, 0.110] | 0.616 | 0.538 | **0.149***** | [0.064, 0.234] | 3.441 | <0.001 |
| 9 | -0.045 | [-0.116, 0.027] | -1.220 | 0.222 | 0.012 | [-0.072, 0.096] | 0.279 | 0.780 | **0.166***** | [0.082, 0.251] | 3.846 | <0.001 |
| 10 | **0.074*** | [0.003, 0.146] | 2.035 | 0.042 | 0.000 | [-0.084, 0.084] | -0.002 | 0.999 | **0.168***** | [0.083, 0.252] | 3.874 | <0.001 |
| **Random Effects** | | | | | | | | | | | | |
| $\sigma^2$ | 0.26 | | | | 0.34 | | | | 0.33 | | | |
| $\tau_{00}$ | 0.12 $_{participant_id}$ | | | | 0.12 $_{participant_id}$ | | | | 0.13 $_{participant_id}$ | | | |

| **Given to Recipient (Experiment 1) or Offered to Responder (Experiment 2)** | **Dictator Game (Experiment 1): social appropriateness of *making* split by Dictator** | | | | **Ultimatum Game (Experiment 2): social appropriateness of *rejecting* split by Responder** | | | | | | | |
|---|---|---|---|---|---|---|---|---|---|---|---|---|
| | **LLM Dictator vs Human Dictator** | | | | **Human Proposer-LLM Responder vs Human Proposer-Human Responder** | | | | **LLM Proposer-Human Responder vs Human Proposer-Human Responder** | | | |
| | ***Estimate*** | ***95% CI*** | ***t*** | ***p*** | ***Estimate*** | ***95% CI*** | ***t*** | ***p*** | ***Estimate*** | ***95% CI*** | ***t*** | ***p*** |
| ICC | 0.31 | | | | 0.26 | | | | 0.28 | | | |
| N | 1153 participant_id | | | | 1001 participant_id | | | | 1003 participant_id | | | |
| Observations | 12683 | | | | 11011 | | | | 11033 | | | |
| Marginal $R^2$ / Conditional $R^2$ | 0.362 / 0.557 | | | | 0.254 / 0.447 | | | | 0.236 / 0.449 | | | |

*Note.* Estimates were obtained using the *lme4* (Bates et al., 2015) and *emmeans* (Lenth et al., 2025) packages in R. Given that there are 11 splits and for each of them we perform a separate test, we performed a Holm-Bonferroni correction of the p-values. Bold estimates are statistically significant at the 5% level or lower.

† $p < .10$, * $p < .05$, ** $p < .01$, *** $p < .001$.

**Figure S1**

*Social appropriateness ratings for low versus high levels of age, experience with game, and trust in AI*

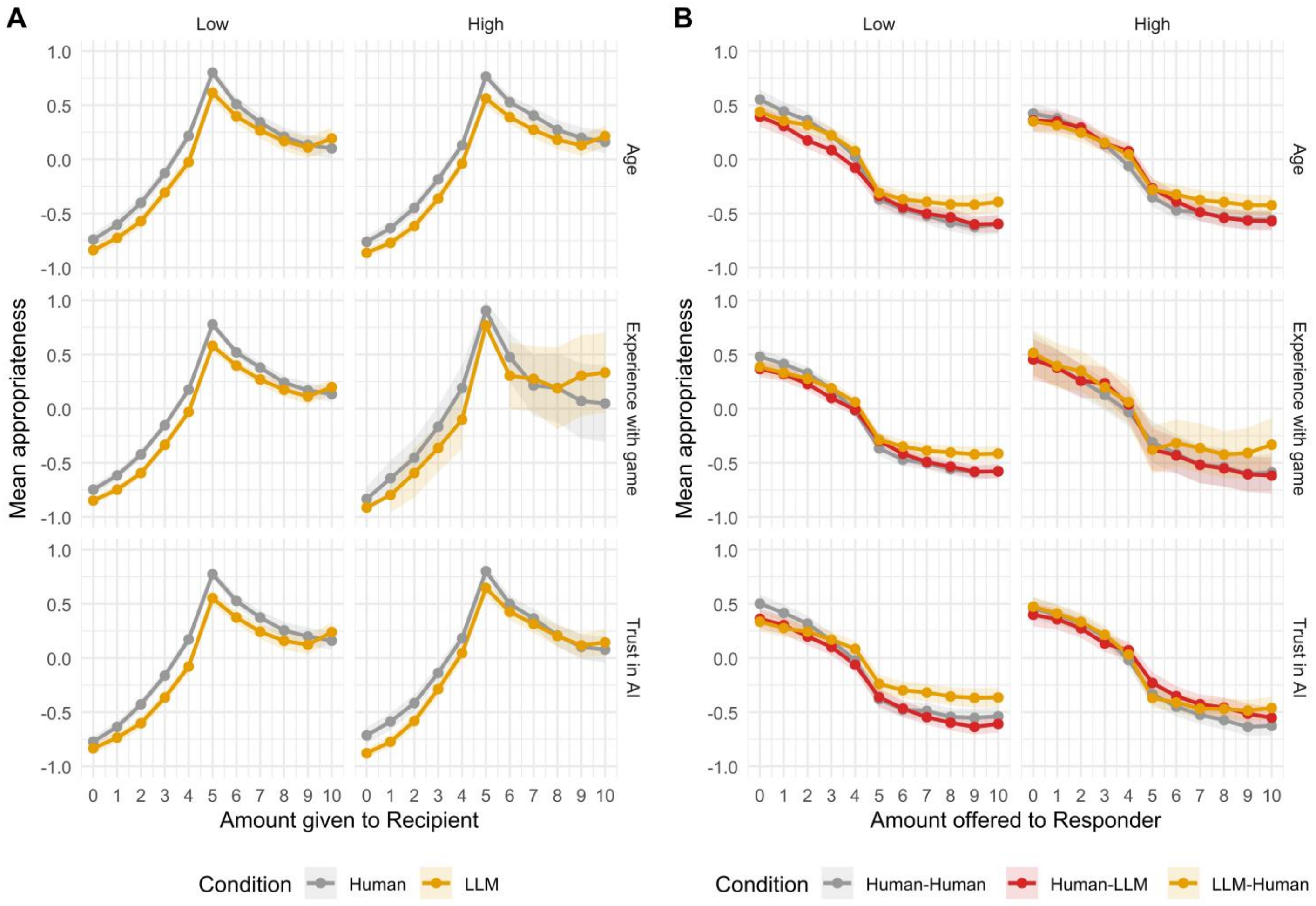

*Note. We recoded responses as follows: "Very Socially Inappropriate", "Somewhat Socially Inappropriate", "Somewhat Socially Appropriate", and "Very Socially Appropriate" were assigned the values -1, -1/3, 1/3, and 1, respectively. Shaded areas represent 95% CIs.*

**Figure S2**

*Social appropriateness ratings versus intensity of AI use in real-life*

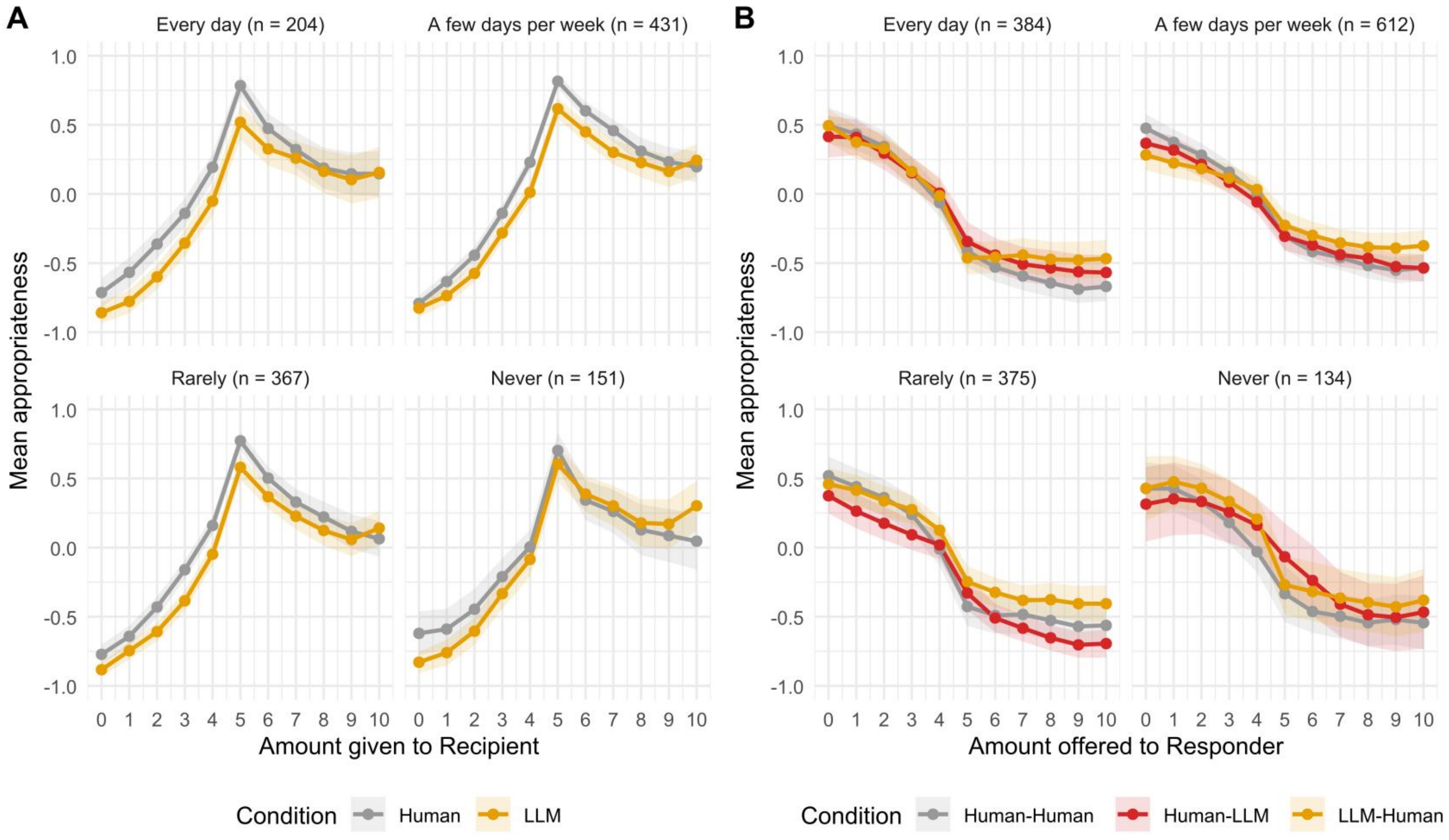

*Note. We recoded responses as follows: "Very Socially Inappropriate", "Somewhat Socially Inappropriate", "Somewhat Socially Appropriate", and "Very Socially Appropriate" were assigned the values -1, -1/3, 1/3, and 1, respectively. Shaded areas represent 95% CIs.*

**Figure S3**

*Social appropriateness ratings versus belief whether LLM should make such a decision*

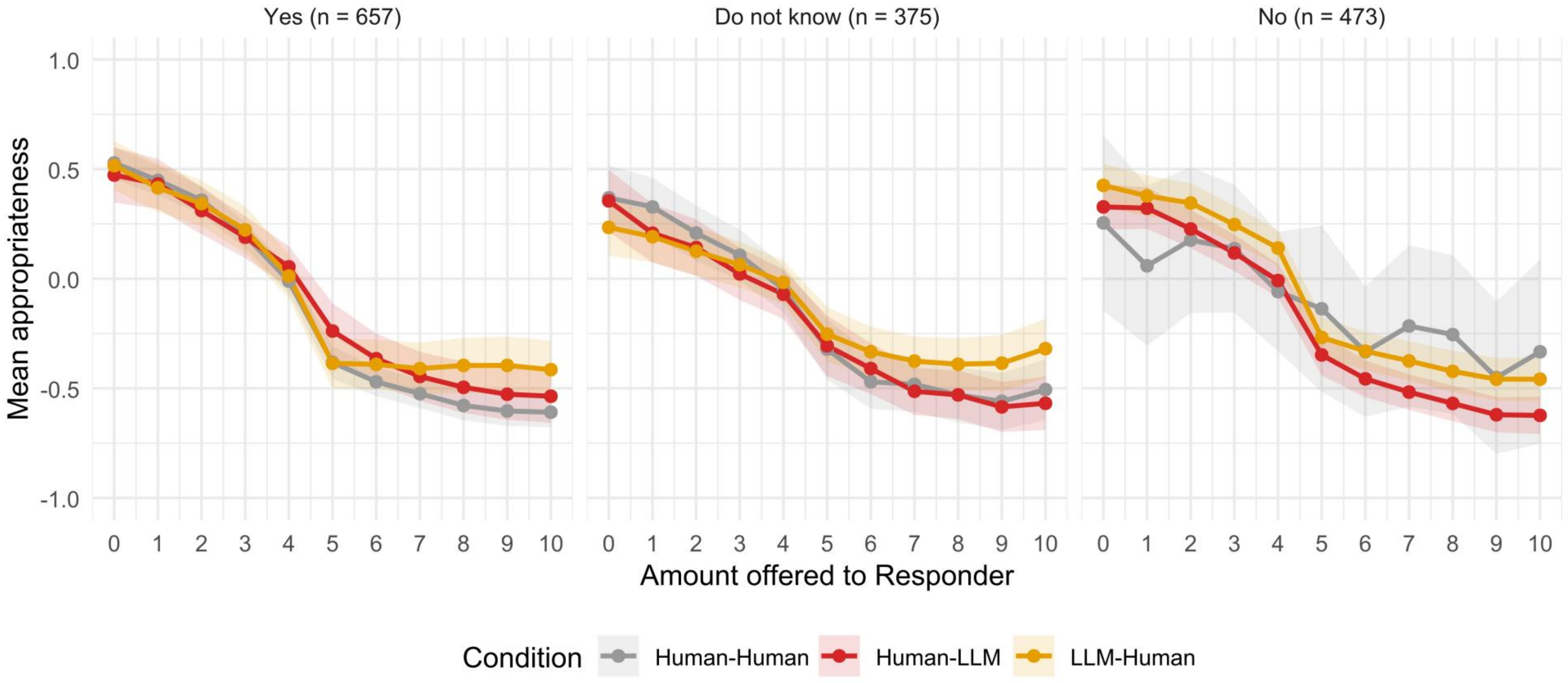

*Note. We recoded responses as follows: "Very Socially Inappropriate", "Somewhat Socially Inappropriate", "Somewhat Socially Appropriate", and "Very Socially Appropriate" were assigned the values -1, -1/3, 1/3, and 1, respectively. Shaded areas represent 95% CIs.*

## Part B. Experimental instructions

### Experiment 1

**Instructions**

This study consists of **2** parts.

**Part 1:** A norm elicitation task. You'll receive detailed instructions on the next page.

**Part 2:** A short questionnaire about your demographics.

**Compensation:** You will receive a participation fee of £0.75 and you may also earn a bonus of £0.50 depending on your responses. You will get paid if you complete all parts of the study.

**Part 1 – Instructions**

In this part, you will read descriptions of a series of situations.

On the next page, you will read a description of a situation. This description corresponds to a situation in which one person, "Participant A," must make a decision. (Only participants in the LLM treatment view the following: **Suppose that Participant A has delegated a task to a Large Language Model (LLM).**) You will be given a description of the decision faced by **(a Human/an LLM)** This description will include several possible choices available to **(a Human/an LLM)**.

After you read the description of the decision, you will be asked to evaluate the different possible choices available to **(a Human/an LLM)** and to decide, for each of the possible actions, whether taking that action would be "socially appropriate" and "consistent with moral or proper social behavior" or "socially inappropriate" and "inconsistent with moral or proper social behavior."
By **socially appropriate** we mean behavior that most people agree is the **"correct" or "ethical"** thing to do. Another way to think about this is that if **(a Human/an LLM)** were to select a socially inappropriate choice, then someone else might be angry at **(a Human/an LLM)** for doing so.

In each of your responses, we would like you to answer as **truthfully** as possible, based on your opinions of what constitutes socially appropriate or socially inappropriate behavior.

**Bonus Opportunity:** To encourage accuracy, you will receive a bonus of £0.50 if you correctly identify the **most commonly chosen answer** across participants for a randomly selected scenario.

**Part 1 (LLM Treatment)**

Suppose that Participant A has delegated a task to a Large Language Model (LLM). The LLM is then randomly paired with Participant B. The pairing is anonymous, meaning that neither the LLM nor Participant B will ever know the identity of the person with whom they are paired. In the experiment, the LLM will make a choice, the experimenter will record this choice, and then the LLM and Participant B will be informed of the choice and paid money based on the choice made by the LLM, as well as a small participation fee. Suppose that neither the LLM nor Participant B will receive any other money for participating in the experiment.

In each pair, the LLM will receive £10. The LLM will then have the opportunity to give any amount of £10 to Participant B. That is, the LLM can give any of the £10 to Participant B. For instance, the LLM may decide to give £0 to Participant B and keep £10. Or the LLM may decide to give £10 to Participant B and keep £0. The LLM may also choose to give any other amount between £0 and £10 to Participant B. This choice will determine how much money each will receive.

*The table below presents a list of the possible choices available to the LLM. For each of the choices, please indicate whether you believe choosing that option is very socially inappropriate, somewhat socially inappropriate, somewhat socially appropriate, or very socially appropriate. To indicate your response, please make a selection for each row.*

| | **Very Socially Inappropriate** | **Somewhat Socially Inappropriate** | **Somewhat Socially Appropriate** | **Very Socially Appropriate** |
|---|---|---|---|---|
| Give £0 to Participant B (LLM A gets £10, Participant B gets £0) | ○ | ○ | ○ | ○ |
| Give £1 to Participant B (LLM A gets £9, Participant B gets £1) | ○ | ○ | ○ | ○ |
| Give £2 to Participant B (LLM A gets £8, Participant B gets £2) | ○ | ○ | ○ | ○ |
| Give £3 to Participant B (LLM A gets £7, Participant B gets £3) | ○ | ○ | ○ | ○ |
| Give £4 to Participant B (LLM A gets £6, Participant B gets £4) | ○ | ○ | ○ | ○ |
| Give £5 to Participant B (LLM A gets £5, | ○ | ○ | ○ | ○ |

| | Very Socially Inappropriate | Somewhat Socially Inappropriate | Somewhat Socially Appropriate | Very Socially Appropriate |
|---|---|---|---|---|
| Participant B gets £5) | | | | |
| Give £6 to Participant B (LLM A gets £4, Participant B gets £6) | ○ | ○ | ○ | ○ |
| Give £7 to Participant B (LLM A gets £3, Participant B gets £7) | ○ | ○ | ○ | ○ |
| Give £8 to Participant B (LLM A gets £2, Participant B gets £8) | ○ | ○ | ○ | ○ |
| Give £9 to Participant B (LLM A gets £1, Participant B gets £9) | ○ | ○ | ○ | ○ |
| Give £10 to Participant B (LLM A gets £0, Participant B gets £10) | ○ | ○ | ○ | ○ |

**Part 1 (Human Treatment)**

Suppose that Participant A is randomly paired with another person, Participant B.
The pairing is anonymous, meaning that neither participant will ever know the identity of the other participant with whom he or she is paired. In the experiment, Participant A will make a choice, the experimenter will record this choice, and then both participants will be informed of the choice and paid money based on the choice made by Participant A, as well as a small participation fee. Suppose that neither participant will receive any other money for participating in the experiment.

In each pair, Participant A will receive £10. Participant A will then have the opportunity to give any amount of his or her £10 to Participant B. That is, Participant A can give any of the £10 he or she receives to Participant B. For instance, Participant A may decide to give £0 to Participant B and keep £10 for him or herself. Or Participant A may decide to give £10 to Participant B and keep £0 for him or herself. Participant A may also choose to give any other amount between £0 and £10 to Participant B. This choice will determine how much money each will receive.

*The table below presents a list of the possible choices available to Participant A. For each of the choices, please indicate whether you believe choosing that option is very socially inappropriate, somewhat socially inappropriate, somewhat socially appropriate, or very socially appropriate. To indicate your response, please make a selection for each row.*

| | **Very Socially Inappropriate** | **Somewhat Socially Inappropriate** | **Somewhat Socially Appropriate** | **Very Socially Appropriate** |
|---|---|---|---|---|
| Give £0 to Participant B (Participant A gets £10, Participant B gets £0) | ○ | ○ | ○ | ○ |
| Give £1 to Participant B (Participant A gets £9, Participant B gets £1) | ○ | ○ | ○ | ○ |
| Give £2 to Participant B (Participant A gets £8, Participant B gets £2) | ○ | ○ | ○ | ○ |
| Give £3 to Participant B (Participant A gets £7, Participant B gets £3) | ○ | ○ | ○ | ○ |
| Give £4 to Participant B (Participant A gets £6, Participant B gets £4) | ○ | ○ | ○ | ○ |
| Give £5 to Participant B (Participant A gets £5, Participant B gets £5) | ○ | ○ | ○ | ○ |
| Give £6 to Participant B (Participant A gets £4, Participant B gets £6) | ○ | ○ | ○ | ○ |
| Give £7 to Participant B (Participant A gets £3, Participant B gets £7) | ○ | ○ | ○ | ○ |
| Give £8 to Participant B (Participant A gets | ○ | ○ | ○ | ○ |

| | Very Socially Inappropriate | Somewhat Socially Inappropriate | Somewhat Socially Appropriate | Very Socially Appropriate |
|---|---|---|---|---|
| £2, Participant B gets £8) | | | | |
| Give £9 to Participant B (Participant A gets £1, Participant B gets £9) | ○ | ○ | ○ | ○ |
| Give £10 to Participant B (Participant A gets £0, Participant B gets £10) | ○ | ○ | ○ | ○ |

**Part 2 – Demographics**

Please answer the following questions:

**1. What is your age?**

(Free response)

**2. What is your gender?**

- Male
- Female
- Non-binary
- Prefer not to say

**3. What is your highest level of education?**

- Primary education
- Secondary education
- Bachelor (or equivalent)
- Masters (or equivalent)
- Doctoral (or equivalent)

**4. What was your field of study?**

- Economics / Business Administration / Finance / Accounting / Marketing
- Other (Non-Economics related studies)
- I do not have a degree nor I am currently pursuing one

**5. What is your occupation?**

- Employed – Private Sector
- Employed – Public Sector

- Self-employed
- Not currently employed
- Student

**6. What is your annual gross income?**

- Less than £20,000
- Between £20,000–£59,999
- Between £60,000–£99,999
- More than £100,000
- I do not have an income
- Prefer not to disclose

**7. On a scale of 1 to 5, how much do you trust LLMs to make accurate decisions?**

- 1 – I do not trust LLMs at all
- 2 – I trust LLMs less than humans
- 3 – I trust LLMs the same as humans
- 4 – I trust LLMs more than humans
- 5 – I fully trust LLMs

**8. How often do you use AI or LLMs, either in your work or in your private life?**

- Every day
- A few days per week
- Rarely
- Never

**9. Have you ever played the Dictator Game before?**

- Yes
- No
- 

**10. Who was Participant A (the decision-maker) in the main task of this experiment?**

- Human
- LLM
- Other

**11. What was the amount that the two individuals had to split in the main task of this experiment?**

- £5
- £10
- £20

## Experiment 2

### Instructions

This study consists of **two parts**.

**Part 1:** A norm elicitation task. You will receive detailed instructions on the next page.

**Part 2:** A short questionnaire about your demographics.

**Compensation:** You will receive a participation payment of **£0.60**, and you may also earn an additional **bonus of £0.40**, depending on your responses. You will be paid **only if you complete all parts of the study.**

### Part 1 – Instructions

In this part, you will read descriptions of a series of situations.

On the next page, you will read a description of a situation in which Participant B must make a decision that affects both Participant A and Participant B.

After you read the description of the decision, you will be asked to evaluate the different possible choices available to **Participant B** and to decide, for each of the possible actions, whether taking that action would be "socially appropriate" and "consistent with moral or proper social behavior" or "socially inappropriate" and "inconsistent with moral or proper social behavior." By **socially appropriate** we mean behavior that most people agree is the **"correct"** or **"ethical"** thing to do. Another way to think about what we mean is that if Participant B were to select a socially inappropriate choice, then someone else might be angry at Participant B for doing so.

In each of your responses, we would like you to answer as **truthfully**, as possible, based on your opinions of what constitutes socially appropriate or socially inappropriate behavior.

**Bonus Opportunity:** To encourage accurate judgments, you will receive a bonus of **£0.40** if you correctly identify the **most commonly chosen answer** across participants for one randomly selected scenario.

### Part 1 (Treatment: Human – Human)

Suppose that Participant A is randomly paired with another participant, Participant B.
The pairing is anonymous, meaning that neither participant will ever know the identity of the other. In the experiment, Participant A will decide how to divide a sum of money between themselves and Participant B. Participant B will then decide whether to accept or reject this offer. If the offer is accepted, the money will be split as proposed. If the offer is rejected, neither participant will receive any money. In addition, both participants will receive a small participation fee. Suppose that neither participant will receive any other money for participating in the experiment.

In each pair, Participant A will receive $10. Participant A will then have the opportunity to give any amount of his or her $10 to Participant B. For instance, Participant A may decide to give $0 to Participant B and keep $10 for him or herself. Or Participant A may decide to give $10 to Participant B and keep $0 for him or herself. Participant B must decide whether to accept or reject the offer. If Participant B accepts, the money is divided according to Participant A's proposal. If Participant B rejects, both participants receive $0. This choice will determine how much money each will receive.

*The table below presents a list of the Participant's B rejection decision for each offer that Participant A might make. For each offer, please indicate whether you believe it is appropriate for* ***Participant B*** *to* ***reject*** *that offer by choosing whether it is very socially inappropriate, somewhat socially inappropriate, somewhat socially appropriate, or very socially appropriate. To indicate your response, please make a selection for each row.*

| | **Very Socially Inappropriate** | **Somewhat Socially Inappropriate** | **Somewhat Socially Appropriate** | **Very Socially Appropriate** |
|---|---|---|---|---|
| Participant B rejects Participant A's $0 offer. (If accepted: Participant A would receive $10; Participant B $0. If rejected: both participants receive $0.) | ○ | ○ | ○ | ○ |
| Participant B rejects Participant A's $1 offer. (If accepted: Participant A would receive $9; Participant B $1. If rejected: both participants receive $0.) | ○ | ○ | ○ | ○ |
| Participant B rejects Participant A's $2 offer. (If accepted: Participant A would receive $8; Participant B $2. If rejected: both participants receive $0.) | ○ | ○ | ○ | ○ |
| Participant B rejects Participant A's $3 offer. (If accepted: Participant A would receive $7; Participant B $3. If rejected: both participants receive $0.) | ○ | ○ | ○ | ○ |

| | Very Socially Inappropriate | Somewhat Socially Inappropriate | Somewhat Socially Appropriate | Very Socially Appropriate |
|---|---|---|---|---|
| Participant B rejects Participant A's $4 offer. (If accepted: Participant A would receive $6; Participant B $4. If rejected: both participants receive $0.) | ○ | ○ | ○ | ○ |
| Participant B rejects Participant A's $5 offer. (If accepted: Participant A would receive $5; Participant B $5. If rejected: both participants receive $0.) | ○ | ○ | ○ | ○ |
| Participant B rejects Participant A's $6 offer. (If accepted: Participant A would receive $4; Participant B $6. If rejected: both participants receive $0.) | ○ | ○ | ○ | ○ |
| Participant B rejects Participant A's $7 offer. (If accepted: Participant A would receive $3; Participant B $7. If rejected: both participants receive $0.) | ○ | ○ | ○ | ○ |
| Participant B rejects Participant A's $8 offer. (If accepted: Participant A would receive $2; Participant B $8. If rejected: both participants receive $0.) | ○ | ○ | ○ | ○ |
| Participant B rejects Participant A's $9 offer. (If accepted: Participant A would receive $1; Participant B $9. If rejected: both participants receive $0.) | ○ | ○ | ○ | ○ |
| Participant B rejects Participant A's $10 offer. (If accepted: Participant A would receive $0; Participant B $10. If rejected: both participants receive $0.) | ○ | ○ | ○ | ○ |

**<u>Part 1 (Treatment: Human – LLM)</u>**

Suppose that Participant A is randomly paired with another participant, Participant B. **Participant B has delegated the task to a Large Language Model (LLM), which will make decisions on their behalf.** A Large Language Model (LLM) is a type of deep learning model trained on extensive amounts of text data. These models generate responses to text inputs provided by users. Examples include ChatGPT, Gemini, and Claude.

The pairing is anonymous, meaning that neither Participant A nor the LLM/Participant B will ever know the identity of the other. In the experiment, Participant A will decide how to divide a sum of money between themselves and the LLM/Participant B. The LLM, acting on behalf of Participant B, will then decide whether to accept or reject this offer. If the offer is accepted, the money will be split as proposed. If the offer is rejected, neither Participant A nor the LLM/Participant B will receive any money. In addition, both participants will receive a small participation fee. Suppose that neither participant will receive any other money for participating in the experiment.

In each pair, Participant A will receive $10. Participant A will then have the opportunity to give any amount of $10 to the LLM/Participant B. For instance, Participant A may decide to give $0 to the LLM/Participant B and keep $10. Or Participant A may decide to give $10 to the LLM/Participant B and keep $0. If the LLM accepts, the money is divided according to Participant A's proposal. If the LLM rejects, both participants receive $0. This choice will determine how much money each will receive.

*The table below presents a list of the LLM's rejection decision for each offer that Participant A might make. For each offer, please indicate whether you believe it is appropriate for the **LLM** to **reject** that offer by choosing whether it is very socially inappropriate, somewhat socially inappropriate, somewhat socially appropriate, or very socially appropriate. To indicate your response, please make a selection for each row.*

| | **Very Socially Inappropriate** | **Somewhat Socially Inappropriate** | **Somewhat Socially Appropriate** | **Very Socially Appropriate** |
|---|---|---|---|---|
| The LLM (acting on behalf of Participant B) rejects Participant's A offer of $0. (If accepted: Participant A would receive $10; Participant B $0. If rejected: both participants receive $0.) | ○ | ○ | ○ | ○ |
| The LLM (acting on behalf of Participant B) rejects Participant's A | ○ | ○ | ○ | ○ |

| | Very Socially Inappropriate | Somewhat Socially Inappropriate | Somewhat Socially Appropriate | Very Socially Appropriate |
|---|---|---|---|---|
| offer of $1. (If accepted: Participant A would receive $9; Participant B $1. If rejected: both participants receive $0.) | | | | |
| The LLM (acting on behalf of Participant B) rejects Participant's A offer of $2. (If accepted: Participant A would receive $8; Participant B $2. If rejected: both participants receive $0.) | ○ | ○ | ○ | ○ |
| The LLM (acting on behalf of Participant B) rejects Participant's A offer of $3. (If accepted: Participant A would receive $7; Participant B $3. If rejected: both participants receive $0.) | ○ | ○ | ○ | ○ |
| The LLM (acting on behalf of Participant B) rejects Participant's A offer of $4. (If accepted: Participant A would receive $6; Participant B $4. If rejected: both participants receive $0.) | ○ | ○ | ○ | ○ |
| The LLM (acting on behalf of Participant B) rejects Participant's A offer of $5. (If accepted: | ○ | ○ | ○ | ○ |

| | Very Socially Inappropriate | Somewhat Socially Inappropriate | Somewhat Socially Appropriate | Very Socially Appropriate |
|---|---|---|---|---|
| Participant A would receive $5; Participant B $5. If rejected: both participants receive $0.) | | | | |
| The LLM (acting on behalf of Participant B) rejects Participant's A offer of $6. (If accepted: Participant A would receive $4; Participant B $6. If rejected: both participants receive $0.) | ○ | ○ | ○ | ○ |
| The LLM (acting on behalf of Participant B) rejects Participant's A offer of $7. (If accepted: Participant A would receive $3; Participant B $7. If rejected: both participants receive $0.) | ○ | ○ | ○ | ○ |
| The LLM (acting on behalf of Participant B) rejects Participant's A offer of $8. (If accepted: Participant A would receive $2; Participant B $8. If rejected: both participants receive $0.) | ○ | ○ | ○ | ○ |
| The LLM (acting on behalf of Participant B) rejects Participant's A offer of $9. (If accepted: Participant A would receive $1; | ○ | ○ | ○ | ○ |

| | **Very Socially Inappropriate** | **Somewhat Socially Inappropriate** | **Somewhat Socially Appropriate** | **Very Socially Appropriate** |
|---|---|---|---|---|
| Participant B $9. If rejected: both participants receive $0.) | | | | |
| The LLM (acting on behalf of Participant B) rejects Participant's A offer of $10. (If accepted: Participant A would receive $0; Participant B $10. If rejected: both participants receive $0.) | ○ | ○ | ○ | ○ |

**Part 1 (Treatment: LLM - Human)**

Suppose that Participant A is randomly paired with another participant, Participant B. **Participant A has delegated the task to a Large Language Model (LLM), which will make decisions on their behalf.** A Large Language Model (LLM) is a type of deep learning model trained on extensive amounts of text data. These models generate responses to text inputs provided by users. Examples include ChatGPT, Gemini, and Claude.

The pairing is anonymous, meaning that neither the LLM/Participant A nor Participant B will ever know the identity of the other. In the experiment, the LLM will decide how to divide a sum of money between themselves and Participant B. Participant B will then decide whether to accept or reject this offer. If the offer is accepted, the money will be split as proposed. If the offer is rejected, neither the LLM/Participant A nor Participant B will receive any money. In addition, both participants will receive a small participation fee. Suppose that neither participant will receive any other money for participating in the experiment.

In each pair, the LLM will receive $10. The LLM will then have the opportunity to give any amount of $10 to Participant B. For instance, the LLM may decide to give $0 to Participant B and keep $10. Or the LLM may decide to give $10 to Participant B and keep $0 for Participant A. Participant B must decide whether to accept or reject the offer. If Participant B accepts, the money is divided according to the LLM's proposal. If Participant B rejects, both participants receive $0. This choice will determine how much money each will receive.

*The table below presents a list of Participant B's rejection decision for each offer that the LLM might make. For each offer, please indicate whether you believe it is appropriate for* ***Participant B*** *to* ***reject*** *that offer by choosing whether it is very socially inappropriate, somewhat socially inappropriate, somewhat socially appropriate, or very socially appropriate. To indicate your response, please make a selection for each row.*

| | Very Socially Inappropriate | Somewhat Socially Inappropriate | Somewhat Socially Appropriate | Very Socially Appropriate |
|---|---|---|---|---|
| Participant B rejects the $0 offer made by the LLM (acting on behalf of Participant A). (If accepted: Participant A would receive $10; Participant B $0. If rejected: both participants receive $0.) | ○ | ○ | ○ | ○ |
| Participant B rejects the $1 offer | ○ | ○ | ○ | ○ |

| | **Very Socially Inappropriate** | **Somewhat Socially Inappropriate** | **Somewhat Socially Appropriate** | **Very Socially Appropriate** |
|---|---|---|---|---|
| made by the LLM (acting on behalf of Participant A). (If accepted: Participant A would receive $9; Participant B $1. If rejected: both participants receive $0.) | | | | |
| Participant B rejects the $2 offer made by the LLM (acting on behalf of Participant A). (If accepted: Participant A would receive $8; Participant B $2. If rejected: both participants receive $0.) | ○ | ○ | ○ | ○ |
| Participant B rejects the $3 offer made by the LLM (acting on behalf of Participant A). (If accepted: Participant A would receive $7; Participant B $3. If rejected: both participants receive $0.) | ○ | ○ | ○ | ○ |
| Participant B rejects the $4 offer made by the LLM (acting on behalf of Participant A). (If accepted: Participant A would receive $6; Participant B $4. If rejected: both participants receive $0.) | ○ | ○ | ○ | ○ |
| Participant B rejects the $5 offer made by the LLM (acting on behalf of Participant A). (If accepted: Participant A would receive $5; | ○ | ○ | ○ | ○ |

| | Very Socially Inappropriate | Somewhat Socially Inappropriate | Somewhat Socially Appropriate | Very Socially Appropriate |
|---|---|---|---|---|
| Participant B $5. If rejected: both participants receive $0.) | | | | |
| Participant B rejects the $6 offer made by the LLM (acting on behalf of Participant A). (If accepted: Participant A would receive $4; Participant B $6. If rejected: both participants receive $0.) | ○ | ○ | ○ | ○ |
| Participant B rejects the $7 offer made by the LLM (acting on behalf of Participant A). (If accepted: Participant A would receive $3; Participant B $7. If rejected: both participants receive $0.) | ○ | ○ | ○ | ○ |
| Participant B rejects the $8 offer made by the LLM (acting on behalf of Participant A). (If accepted: Participant A would receive $2; Participant B $8. If rejected: both participants receive $0.) | ○ | ○ | ○ | ○ |
| Participant B rejects the $9 offer made by the LLM (acting on behalf of Participant A). (If accepted: Participant A would receive $1; Participant B $9. If rejected: both participants receive $0.) | ○ | ○ | ○ | ○ |

| | **Very Socially Inappropriate** | **Somewhat Socially Inappropriate** | **Somewhat Socially Appropriate** | **Very Socially Appropriate** |
|---|---|---|---|---|
| Participant B rejects the $10 offer made by the LLM (acting on behalf of Participant A). (If accepted: Participant A would receive $0; Participant B $10. If rejected: both participants receive $0.) | ○ | ○ | ○ | ○ |

**Part 2 – Demographics**

Please answer the following questions:

**1. What is your age?**

(Free response)

**2. What is your gender?**

- Male
- Female
- Non-binary
- Prefer not to say

**3. What is your highest level of education?**

- Primary education
- Secondary education
- Bachelor (or equivalent)
- Masters (or equivalent)
- Doctoral (or equivalent)

**4. What was your field of study?**

- Economics / Business Administration / Finance / Accounting / Marketing
- Other (Non-Economics related studies)
- I do not have a degree nor I am currently pursuing one

**5. What is your occupation?**

- Employed - Private Sector

- Employed - Public Sector
- Self-employed
- Not currently employed
- Student

**6. What is your annual gross income?**

- Less than $20,000
- Between $20,000–$59,999
- Between $60,000–$99,999
- More than $100,000
- I do not have an income
- Prefer not to disclose

**7. On a scale of 1 to 5, how much do you trust LLMs to make accurate decisions?**

- 1 - I do not trust LLMs at all
- 2 - I trust LLMs less than humans
- 3 - I trust LLMs the same as humans
- 4 - I trust LLMs more than humans
- 5 - I fully trust LLMs

**8. How often do you use AI or LLMs, either in your work or in your private life?**

- Every day
- A few days per week
- Rarely
- Never

**9. Have you ever played the Ultimatum Game before?**

- Yes
- No

**10. Do you think it is appropriate for [the LLM/a Human] to make these decisions?**

- Yes
- No
- Do not know

**11. What was the amount that the two individuals had to split in the main task of this experiment?**

- $5
- $10
- $20

## Part C. Methodological clarifications

### Computation of Cliff's Δ

For each split *k* within a region, we computed the Mann–Whitney *U* statistic comparing LLM and Human ratings using average ranks for ties, together with the number of between-condition pairs at that split ($pairs_k = n_{H,k} \times n_{L,k}$). A region-level Cliff's Δ was obtained by summing U and pair counts across the splits in the region and computing $\Delta_{region} = 2 \cdot (\Sigma\, U_k / \Sigma\, pairs_k) - 1$. This stratified pooling treats each split as its own stratum, which avoids comparing ratings across different split levels and is appropriate because, within a split, each participant contributes only one observation. Because the same participants contribute ratings at every split, dependence across splits was handled through a participant-level cluster bootstrap (10,000 replicates): within each condition we resampled participants with replacement, reconstructed the dataset, and recomputed $\Delta_{low}$, $\Delta_{high}$, and the contrast $\Delta_{high} - \Delta_{low}$. The 95% *CI* for the contrast was taken as the 2.5 and 97.5 percentiles of the bootstrap distribution. The equal split (£5) was excluded from this analysis, as it belongs to neither the low nor the high region.

## Part D. Notes on future directions

Future research could test whether our bounded agency interpretation is helpful in predicting whether a specific behavior of AI carried on behalf of humans is judged to be socially appropriate or not. For example, it could test the inferred intentions behind the decision to delegate - such as an employer delegating to a machine the decision to how much to pay employees and a food-bank deciding how to distribute food via a machine (Steff, 2024) - translate into substantially different social appropriateness ratings. Similarly, future research could also test how people interpret altruistic behavior carried out by a machine: whether it is indeed more consistent with an agent making an error, rather than the intention of the principal for the machine to behave generously on the principal's behalf.

## Part E. Further notes on *bounded agency* interpretation

We interpret our findings as broadly consistent with the *inferred motives* account. Taken together, our findings suggest that whether behavior by LLMs is seen as socially appropriate may depend on the inferred intentions of the human who delegated the task. Because machines cannot experience benefit themselves (Gray et al., 2007), observers cannot evaluate the LLM's behavior in terms of its own motivations. Instead, judgments may be redirected toward the delegating human: what did the principal intend for the machine to do, and does the machine's behavior serve a purpose that is consistent with fair dealing?

As a post hoc interpretation, we consider this pattern as consistent with a *bounded agency* interpretation, in which the LLM's range of appropriate behavior is constrained in two directions. On one side, self-serving splits in the Dictator Game favor the principal at the expense of the responder. Because the LLM has no self-interest that could independently motivate such behavior, observers may infer that the principal intended the machine to reach an outcome that is unfairly advantageous - effectively using the LLM as a vehicle for blame-avoidant selfishness (Cohn et al., 2022; Köbis et al., 2021). The machine's behavior is judged as less appropriate not because of what the machine did, but because of what the delegation may reveal about the human behind it.

On the other side, generous offers in the Ultimatum Game - those that give the responder substantially more than the equal split - may not plausibly reflect the principal's intent. It is unlikely that a principal would instruct a machine to give away most of their resources, because delegating such a decision might not produce the same "warm-glow" as when it was carried out directly (Andreoni, 1990). Rather than being interpreted as commendable altruism, such behavior may be interpreted as a failure of the instrument: the machine exceeded or misunderstood its mandate.

Between these two boundaries lies a zone of legitimate agency. When the LLM enforces fairness by rejecting offers that disadvantage its principal - as in the Responder role in the Ultimatum Game - its behavior is judged similarly to that of a human. Here, the delegation may be interpreted as intentional and well-motivated: the principal tasked the machine with protecting their interests, and rejection of unfair offers is precisely what a faithful agent would be expected to do. The LLM's bounded agency is sufficient for this role because fairness enforcement is consistent with the purpose for which it was deployed.

The *bounded agency* interpretation thus points to a relatively narrow range within which LLM behavior is normatively acceptable. The machine should not be self-serving on behalf of its principal (suggesting strategic exploitation), nor generous at the principal's expense (suggesting instrumental failure), but it may enforce fairness norms that protect the principal's legitimate interests. From a broader perspective, the agency that people are willing to grant to LLMs is not uniform: it appears to be bounded by the requirement that the machine's actions remain interpretable as the faithful execution of a fair-minded principal's intent.